\documentclass{article}
\usepackage{amsmath}
\usepackage{amsfonts}
\usepackage[latin1]{inputenc}
   
\newcommand{\di}{\displaystyle}

\newcommand{\R}{\mathbb  R}

\newcommand{\N}{\mathbb  N}
\newcommand{\1}{\mathbb  I}
\newcommand{\la}{\langle}
\newcommand{\ra}{\rangle}

\newtheorem{lemma}{Lemma}[section]
\newtheorem{theorem}[lemma]{Theorem}
\newtheorem{proposition}[lemma]{Proposition}
\newtheorem{corollary}[lemma]{Corollary}
\newtheorem{remark}[lemma]{Remark}

\newtheorem{definition}[lemma]{Definition}

\def\noi{\noindent}

\def\beq{\begin{equation}}   \def\eeq{\end{equation}}
\def\bea{\begin{eqnarray}}  \def\eea{\end{eqnarray}}
 
\def\noi{\noindent}

\newcommand\mysection{\setcounter{equation}{0}\section}
\renewcommand{\theequation}{\thesection.\arabic{equation}}
\newcounter{hran} \renewcommand{\thehran}{\thesection.\arabic{hran}}

\def\bmini{\setcounter{hran}{\value{equation}}
    \refstepcounter{hran}\setcounter{equation}{0}
    \renewcommand{\theequation}{\thehran\alph{equation}}\begin{eqnarray}}

\def\bminiG#1{\setcounter{hran}{\value{equation}}
\refstepcounter{hran}\setcounter{equation}{-1}
\renewcommand{\theequation}{\thehran\alph{equation}}
\refstepcounter{equation}\label{#1}\begin{eqnarray}}

\begin{document}

\title {Time Evolution of States for Open Quantum Systems.  \\
The quadratic case\footnotemark\footnotetext{This work  was supported by the French Agence Nationale de la Recherche,
 NOVESOL project, ANR 2011, BS0101901}}
\author{ Didier Robert }

\vskip 1 truecm
\date{}
\maketitle

\begin{abstract}
Our  main goal  in this paper is to extend  to any  system of coupled quadratic Hamiltonians 
some properties  known for  systems of quantum  harmonic  oscillators   related with the  Brownian Quantum  Motion model. 
In  a first part  we get  a  rather general  formula for the purity (or the linear entropy) in a short time approximation. 
For this formula the quadratic assumption is not necessary, more general Hamiltonians can be considered.\\
In a second part we establish a  master equation (or a Fokker-Planck type equation)
 for the time evolution  of the reduced matrix density  for  bilinearly coupled quadratic Hamiltonians.
 The  Hamiltonians and  the bilinear coupling can  be time dependent.   \\
 Moreover we give   an explicit  formula  for the solution of this  master equation so that
  the time evolution of the reduced density at  time $t$ is  written as a convolution integral  for the reduced 
  density at  initial time $t_0=0$,  with a Gaussian kernel,  for $0 \leq t < t_c$ where $t_c\in ]0, \infty]$ is  a critical time. Reversibility is lost for $t \geq  t_c$.
\end{abstract}

\pagestyle{myheadings}

\mysection{Introduction }
The general setting  considered here is  a quantum system  $(S)$  interacting with an environment  $(E)$.
The total system  $(S)\cup (E)$ is  supposed to be   an isolated  quantum system and we are interested in dynamical
 properties of  $(S)$ alone,   which is an open system because of  its interactions with $(E)$.
In particular during  the time evolution the energy of  $(S)$  is not preserved and  its evolution is not determined by a Schr\"odinger or Liouville-von Neumann equation unlike for  the total system $(S)\cup (E)$.

\noi
The Hilbert space of the total system $(S)\cup(E)$ is   the tensor product 
  ${\cal H} ={\cal H}_S\otimes{\cal H}_E $\footnote{ When the spaces are infinite dimensional we 
  mean  here that  ${\cal H}_S\otimes{\cal H}_E$ is the Hilbert tensor product (completion
of the algebraic tensor product). }  and its Hamiltonian
$\hat H$  is decomposed as follows:
\beq
\hat H = \underbrace{\hat H_S\otimes\1 + \1\otimes\hat H_E}_{\hat H_0} + \hat H_I = \hat H_0 + \hat V
\eeq
$(S)\cup(E)$ is isolated by assumption and its evolution obeys the Schr\"odinger (or Liouville-von Neumann) equation with the 
Hamiltonian $\hat H$.\\
All Hamiltonians   here are self -adjoint operators  on their  natural domain,  
 $\hat H$ and $\hat H_I=\hat V$ are defined in ${\cal H}$, $\hat H_S$ in  $\cal H_S$, 
 $\hat H_E$ in ${\cal H}_E$.  Moreover most of the results stated here are valid when the Hamiltonians are time dependent, assuming that their  propagators  exist as unitary operators in the  corresponding Hilbert space.\\
  Quantum observables are denoted with a  hat accent,
  the  corresponding classical observables (also named  Wigner function or Weyl-symbols)  are written    by erasing the hat.\\
  We don't give here more details concerning the domains, these will be clear in the applications.\\
We assume that  the interacting  potential $\hat V$ has the following form
\beq\label{inter1}
 \hat V = \sum_{1\leq j\leq m} \hat S_j\otimes\hat E_j
\eeq
 where $\hat S_j$ and $\hat E_j$ are self-adjoint operators in ${\cal H}_S$, ${\cal H}_E$ respectively.
 
  Recall that a density matrix $\hat\rho$ is a positive class trace operator with trace one (${\rm tr}\hat\rho=1$). 
  $\hat\rho$ is a  state of the total system $(S)\cup(E)$ in the Hilbert space ${\cal H}$.
  The time evolution of $\hat\rho$  obeys
  the following Liouville-von Neumann equation:
  \beq\label{LN}
  \dot{\hat\rho} =  i^{-1}[\hat H, \hat\rho],\;\; \hat\rho: t\mapsto\hat\rho(t)
  \eeq
  Time derivatives are denoted by a dot, $[\cdot,\cdot ]$ denotes the commutator of two observables.\\
We assume that for $t=0$ the system and the environment are decoupled :
\beq
\hat\rho(0) =\hat\rho_S(0)\otimes\hat\rho_E(0),\; \hat\rho_{S,E}(0) \;{\rm are \;  density \; matrices\; in}\; {\cal H}_{S,E}
\eeq
  and the state $\hat\rho_S(0)$ of the system is pure i.e is an orthogonal projector on a unit vector $\psi$ of ${\cal H}_S$.
  A non pure state will be called a mixed state. A density matrix $\hat\rho_S$ is a mixed state if and only if $\hat\rho_S$ has an eigenvalue $\lambda$, such that  $0 < \lambda<1$.\\
If  $\hat\rho$  is a pure state  then $\hat\rho =\Pi_\psi$, $\psi\in{\cal H}$, $\Vert\psi\Vert =1$ where 
 $\Pi_\psi(\eta) = \la\psi, \eta\ra\psi$.\\    
  
  \noi
  The general problem for open systems  is to describe the evolution   $\hat\rho_S(t)$ of the density matrix   of the system $(S)$.
  In particular an important physical  question is: when  the state $\hat\rho_S(t)$ is pure or mixed?  
  
  \noi
  To decide if a state is pure or not we consider the purity function
  $$
  p_{ur}(t) = {\rm tr}_S(\hat\rho_S(t)^2)
  $$
  The function ${\rm S}_\ell(t) = 1 -p_{ur}(t) $ is called the linear entropy. The Boltzman-von Neumann entropy is
  $ {\rm S}_{BN} = -{\rm tr }_S(\hat\rho_S\log\hat\rho_S)$.\\
We clearly have $ 0\leq p_{ur}(t)\leq 1 $.  
  It is not difficult to see that $\rho_S(t)$ is pure  if and only if $p_{ur}(t) = 1 $.\\
 With our assumption, $0$ is a maximum for $p_{ur}$  so we have $\ddot{p}_{ur}(0) \leq 0$ and 
 if $\ddot{p}_{ur}(0) <  0$ then for $0<t<\varepsilon$,   $\hat\rho_S(t)$ {\rm } is a mixed state  for $\varepsilon$ small enough.
 \\
 For isolated  (closed)  systems, if the initial state is pure then it stays pure at every time, as it can be easily seen using  equation (\ref{LN}).\\
 So our first step  in this paper  is to compute $\ddot{p}_{ur}(0)$. In particular for a large class of  models, including the Quantum Brownian Motion, 
  we shall prove that $\ddot{p}_{ur}(0)<0$ so  for these models the state of the system becomes  very quickly decoherent and entangled with the environment.

      \noi
 Our second goal is to compute  the time evolution  $\hat\rho_S(t)$ of the state of the system. We know that  $\hat\rho_S(t)$ does not satisfy a   Liouville- von Neumann equation because the system is open. \\
 We shall prove that for any time dependent quadratic systems,  $\hat\rho_S(t)$ obeys an exact  "master equation" similar to a Fokker-Planck  type equation,  with time dependent coefficients,
 and from this equation we can get an explicit formula for $\hat\rho_S(t)$.\\
   This could be a first step for more general  systems
   in the semi-classical r\'egime .\\
  
  \noi
   Let us  recall  a mathematical definition for $\rho_S(t)$ and related properties and notations. \\
   This can be done by introducing partial trace 
  (or relative trace) over the environment for a state $\hat\rho$ of the global system.
  \begin{definition}\label{partr}
  Let $\hat A$ be a class-trace operator in ${\cal H}_S\otimes{\cal H}_E$. We denote ${\rm tr}_E\hat A$ the unique trace-class operator  on ${\cal H}_S$ satisfying, 
  for every  bounded operator $\hat B$ on ${\cal H}_S$,  
  \beq\label{partra}
  {\rm tr}_S\left(({\rm tr}_E\hat A)\hat B\right) = {\rm tr}\left(\hat A(\hat B\otimes\1_E)\right)
  \eeq
  \end{definition}
 We have  denoted  ``${\rm tr}$"  the trace in the total  space ${\cal H}$, ${\rm tr}_{S,E}$ the trace in ${\cal H}_{S,E}$ respectively.\\
 Of course we have the "Fubini property": 
 ${\rm tr}_S({\rm tr}_E\hat A ) = {\rm tr}\hat A$. Moreover ${\rm tr}_E(\hat B\otimes \hat C) = \hat B\cdot({\rm tr}_E\hat C)$ if $\hat B$ and $\hat C$ are trace class operators in ${\cal H}_S$ and ${\cal H}_E$ respectively.
 
 \noi
If $\hat\rho$ is  a density matrix  in ${\cal H}$ then  ${\rm tr}_{E}\hat\rho$ is a density matrix in ${\cal H}_S$,
 called the reduced density matrix or the reduced state.\\
From (\ref{partra}) we  can  compute the matrix element of ${\rm tr}_E\hat A$ by   the formula
$$
\la\psi, ({\rm tr}_E\hat A)\varphi\ra = {\rm tr}\left(\hat A(\Pi_{\psi,\varphi}\otimes\1_E)\right),\;\;\forall \psi, \varphi\in{\cal H}_S.
$$
The partial trace was introduced in quantum mechanics to explain  quantum phenomena   like  {\em entanglement}
 and {\em decoherence}  (see \cite{nich, schl} for more details).\\
Let  $\Psi\in{\cal H}_S\otimes{\cal H}_E$, $\di{\Psi = \sum_{1\leq j\leq N}\psi_j\otimes\eta_j}$. $\Psi$ is a pure state
in ${\cal H} = {\cal H}_S\otimes{\cal H}_E$.  Let us compute the partial trace of $\Pi_\Psi$ in ${\cal H}_E$. Applying Definition \ref{partr}
we easily get
\beq
{\rm tr}_E(\Pi_\Psi) = \sum_{1\leq j,k\leq N}\la\eta_j,\eta_k\ra\Pi_{\psi_j,\psi_k}
\eeq
where $\Pi_{\psi_j,\psi_k}$ is the rank one operator in ${\cal H}_S$ : $\Pi_{\psi_j,\psi_k}(\varphi) = \la\psi_j,\varphi\ra\psi_k$.\\
Decoherence means in particular  that  there exists a (non-orthonormal)  basis $\{\eta_j\}$ such that    $\vert\la\eta_j,\eta_k\ra\vert $  becomes very small for $j\neq k$, so that ${\rm tr}_E(\Pi_\Psi) $
 is very close to $\di{\sum_{1\leq j\leq N}p_j\Pi_{\psi_j}}$ ($p_j=\la\eta_j,\eta_j\ra$) hence  quantum interferences for the system $(S)$  are lost.

  \noi
  Formula (\ref{partra}) is an operator version of Fubini integration theorem as we can see in the Weyl quantization setting.\\
  Let ${\cal H}_S =L^2(\R^d)$ ,\;  ${\cal H}_E =L^2(\R^N)$. 
  Denote $z=(x,\xi)\in\R^{2d}$, $u=(y,\eta)\in\R^{2N}$, $A=\sigma_w\hat A$ the  Weyl symbol
   of $\hat A$  in  the Schwartz space ${\cal S}(\R^{2(d+N)})$ (see \cite{corob} for more details)  then the Weyl symbol of ${\rm tr}_E\hat A$ 
   is 
\beq
   \sigma_w({\rm tr}_E\hat A)(z) = (2\pi)^{-N}\int_{\R^{2N}}A(z,u)du.
\eeq

 Our main applications here concern the Weyl-quadratic case where :
 \begin{itemize}
 \item $H_S$, $ H_E$ are quadratic Hamiltonians (generalized harmonic  oscillators)  respectively in the phase spaces $\R^{2d}$, $\R^{2N}$.
 \item $\hat H_S$, $\hat H_E$ are quantum Hamiltonians (Weyl quantization of $H_S$ respectively $H_E$)
 in $L^2(\R^d) $ respectively $L^2(\R^N)$.
 \item $\hat\rho_S$ is the projector on $\psi \in {\cal S}(\R^d)$.
 \item $\rho_E$ is a Gaussian   with  0 means.
 \item $S_j$ resp. $E_j$ are linear forms on $\R^{2d}$ resp. $\R^{2N}$ (bilinear coupling).
  \end{itemize}
  In applications  one consider  thermal equilibrium states for environment: 
  $\hat\rho_E =  Z(\beta)^{-1}{\rm e}^{-\beta\hat K_E}$
   where $K_E$ is a quadratic form, positive-definite and $\beta =\frac{1}{T}$, $T>0$ is the temperature,
    $Z(\beta) = {\rm tr}({\rm e}^{-\beta\hat K_E})$.\\
In this  setting it is possible to find the exact  time dependent master equation satisfied  by $\hat\rho_S(t)$,  
 as we shall see in the  second part of this paper ($\hat\rho_E$  can be any environment state)  and to solve this equation   when the classical dynamics of the total system $(S)\cup(E)$ is known.
 
 Let us remark that there exists a general formula, called Kraus formula,   giving  the time evolution of  the 
  reduced system:
 $\hat\rho_S(t) = {\rm tr}_E(U(t)\hat\rho(0)U^*(t))$ where $\hat\rho(0) = \hat\rho_S(0)\otimes\hat\rho_E(0)$  and  $U(t) = {\rm e}^{-it\hat H}$
  (see \cite{schl, brpe}):
 \beq\label{kraus}
 \hat\rho_S(t) =    \sum_{j,\ell} \hat K_{j,\ell}(t)\hat\rho_S(0)\hat K_{j,\ell}(t)^*
 \eeq
 where the Kraus operators $\hat K_{j,\ell}(t)$ depends on $\hat\rho_E(0$ and $U(t)$.
 But  the operators  $\hat K_{j,\ell}(t)$ are not explicitly known,  so formula (\ref{kraus}) is not easy to use  to get properties
 of  $\hat\rho_S(t)$.


 \section { A general computation}
 Our goal in this section  is to  give a local formula  in time for the purity when the system $(S)$ is in a pure state.\\
  
 It is easier here  to consider the interaction formulation of quantum mechanics to "eliminate" the   "free" (non-interacting) evolution:
  $U_0(t) = {\rm e}^{-it\hat H_0}$.  In this section we   consider  the general  setting described in the introduction.
  In particular we assume that  the interaction Hamiltonian $\hat V$  has  the shape (\ref{inter1}). 
 As it is well known the time evolution ({\ref{LN})  is given by the Schr\"odinger evolution
 $$
 \hat\rho(t) = U(t)\hat\rho(0)U(t)^*,\;\; U(t) = {\rm e}^{-it\hat H}.
 $$
 For simplicity we assume here that all Hamiltonians are time independent but the results are also valid if
 $H, S_j, E_j$ are  time dependent. In this case we have  $U(t) = U(t,0)$ where $0$ is the initial time.\\
 Some technical assumptions are necessary to get rigorous results.  \\
 Let us denote ${\mathfrak S}^p({\cal H})$ the Schatten class of linear operators in the Hilbert ${\cal H}$ such that
 $\di{\sum_{j\geq 1}\lambda_j(\vert\hat A\vert)^p <+\infty}$, where $\vert\hat A\vert =(\hat A^*A)^{1/2}$
 and $\{\lambda_j(\vert\hat A\vert)\}$ is the eigenvalues series  of $\vert\hat A\vert $.
 For properties  of Schatten classes see \cite{gokr}.
 In particular ${\mathfrak S}^1({\cal H})$ is the trace class space and ${\mathfrak S}^2({\cal H})$ is the Hilbert-Schmidt
 space.
 Our general assumptions are the following.\\
 $(GSE)$. There exist self-adjoint operators in ${\cal H}$, $K_0$ and $K_1$ such that $K_1$  is bounded, 
 $\hat K_0$ is positive with a bounded inverse $\hat K_0^{-1}$  in the Hilbert-Schmidt space ${\mathfrak S}^2({\cal H})$
 and such that
 $$
 \hat\rho(0) = \hat K_0^{-1}\hat K_1^2\hat K_0^{-1}.
 $$
 \noi
 $(Ham)_k$.   For all $j\leq k$ ($k\in\N$),   
 $\hat H^j\hat K_0^{-1}$ is in the Hilbert-Schmidt space ${\mathfrak S}^2({\cal H})$. 
 \begin{lemma}
If  assumptions  $(GSE)$ and  $(Ham)_k$  are satisfied then   $t\mapsto\hat\rho(t)$
 is $C^k$ smooth  in time $t$,  from $\R$ into the trace class Banach space  ${\mathfrak S}^1({\cal H})$
 \end{lemma}
 The proof is standard and is left to the reader.

  It is easier here  to consider the interaction formulation of quantum mechanics to "eliminate" the   "free" (non-interacting) evolution:
  $U_0(t) = {\rm e}^{-it\hat H_0}$
 Introduce the inter-acting evolution 
 $$
 \hat\rho^{(I)}(t) = U_0(t)^*\hat\rho(t)U_0(t),
 $$
 (\ref{LN}) becomes
 \beq\label{LNI}
  \dot{\hat\rho}^{(I)} =  i^{-1}[V(t), \hat\rho^{(I)}],
 \eeq
 where 
 $$
 \hat V(t) = U_0(t)^*\hat VU_0(t) = \sum_{1\leq j\leq m}S_j(t)\otimes E_j(t),
 $$
 with $\hat S_j(t) = {\rm e}^{it\hat H_S} \hat S_j{\rm e}^{-it\hat H_S}$  and  
 $\hat E_j(t) = {\rm e}^{itH_E} \hat E_j{\rm e}^{-itH_E}$.
 The usefull  fact here is that the shape of the interaction $\hat V(t)$ is the same for all $t\in\R$ (\ref{inter1}).\\
 Let us denote by ${\rm Dom}(\hat K)$ the domain of the self-adjoint operator $\hat K$ and 
 introduce the  following  assumptions on $\hat\rho_S(0)$ and $\hat\rho_E(0)$.\\
 $(S)$  $\hat\rho_S(0)$ is a pure state:  $\hat\rho_S(0) =\Pi_{\psi_S}$ with
  $\di{\psi_S\in\bigcap_{1\leq j\leq m}{\rm Dom}(\hat S_j)}$ and  
  for any $1\leq j'\leq m$, $\di{\hat S_{j'}\psi_S\in \bigcap_{1\leq j\leq m}{\rm Dom}(\hat S_j)}$.\\
  $(E)$ For any $\ell, \ell'$ in $\{1, \cdots, m\}$, the following operators are   trace class in the Hilbert space
  ${\cal H}_E$: $\hat E_j\hat\rho_E(0)$, $\hat E_j\hat\rho_E(0)\hat E_{j'}$, $\hat E_{j'}\hat E_j\hat\rho_E(0)$.

 \noi 
 We  easily see  that 
 $$
 \hat\rho_S^{(I)}(t) := {\rm tr}_E(\hat\rho^{(I)}(t)) = {\rm e}^{it\hat H_S}\hat\rho_S(t) {\rm e}^{-it\hat H_S}.
$$
We want  to consider the purity  $p_{ur}(t) = {\rm tr}_S(\hat\rho_S^I(t)^2)$.
 We shall use now the simpler notation $\hat r(t) = \hat\rho_S^{(I)}(t)$.\\
 Using cyclicity of the trace and assumption $(Ham)_2$  we have
\bea
 \dot p_{ur}(t) &=& 2 {\rm tr}_S(\hat r(t)\dot{\hat r}(t)) \\
 \ddot p_{ur}(t)
  &= & 2\left({\rm tr}_S(\hat r(t)\ddot{\hat r}(t)) + {\rm tr}_S(\dot{\hat r}(t))^2\right)
\eea
We have, at time $t=0$, 
\bea
\dot{\hat r}(0) &=& i^{-1}{\rm tr}_E[\hat V, \hat\rho_S\otimes\hat\rho_E], \\
\ddot{\hat r}(0) &=& i^{-1}{\rm tr}_E[\dot{\hat V}, \hat\rho_S\otimes\hat\rho_E] - {\rm tr}_E[\hat V, [\hat V, \hat\rho_S\otimes\hat\rho_E]].
\eea
\begin{lemma}\label{anulenv} 
 Assume that conditions  $(GSE)$, $(Ham)_2$ for  $\hat H$ and $\hat H_0$, $(S)$  and $(E)$ are satisfied 
 and that ${\rm tr}_E(\hat\rho_E(0)\hat E_j) = 0$ for every $1\leq j\leq m$. Then we have  
 $ \dot{\hat r}(0)  = 0$.
 \end{lemma}

 {\em Proof.}  
For simplicity we denote $\hat\rho_E=\hat\rho_E(0)$,  $\hat\rho_S=\hat\rho_S(0)$. Using the splitting  assumption on $\hat V$ we have 
\bea
 [\hat V, \hat\rho_S\otimes\hat\rho_E] &=&\big[\sum_{1\leq j\leq m}\hat S_j\otimes\hat E_j, \hat\rho_S\otimes\hat\rho_E\big]\\
&=& \sum_{1\leq j\leq m}\hat S_j\hat\rho_S\otimes \hat E_j\hat\rho_E -\hat\rho_S\hat S_j\otimes \hat\rho_E\hat E_j 
\eea
So taking the $E$-trace we have
$$
{\rm tr}_E( [\hat V, \hat\rho_S\otimes\hat\rho_E]) = \sum_{1\leq j\leq m}[\hat S_j, \hat\rho_S]{\rm tr}_E(\hat\rho_E\hat E_j)
$$
$\square$\\
\begin{lemma}
Under the assumptions of Lemma \ref{anulenv} we have
\beq\label{nulsq}
{\rm tr}_S\left({\rm tr}_E[\dot{\hat V}, \hat\rho_S\otimes\hat\rho_E]\hat\rho_S\right) =0
\eeq
\end{lemma}
{\em Proof}. 
This follows from the definition of the relative trace ${\rm tr}_E$ and cyclicity of the trace.\\
We have
$$  
{\rm tr}_S\left({\rm tr}_E[\dot{\hat V}, \hat\rho_S\otimes\hat\rho_E]\hat\rho_S\right)  =
{\rm tr}\left([\dot{\hat V}, \hat\rho_S\otimes\hat\rho_E]\hat\rho_S\otimes 1\right) 
$$
This is 0 because $\hat\rho_S\otimes\hat\rho_E$ and $\hat\rho_S\otimes 1$ commute.
$\square$

\noi
Finally, using the identity 
$$
[\hat V,[\hat V,\hat\rho]] = \hat V^2\hat\rho + \hat\rho\hat V^2 -2\hat V\hat\rho\hat V
$$
 with the same argument as in Lemma \ref{anulenv} we get the following proposition:
\begin{proposition} Under the  conditions  of Lemma \ref{anulenv} we have
\beq\label{repur}
 \ddot p_{ur}(0) = -4{\rm tr}\left(\hat V\hat\rho\hat V( \hat\rho^\perp_S\otimes\1)\right)
\eeq
with  $\hat\rho = \hat\rho_S\otimes\hat\rho_E$ and $\hat\rho^\perp_S = \1-\hat\rho_S$ where $\rho_{S,E}=\rho_{S,E}(0)$.\\
In particular if there exist $u, v\in{\rm Range}( \hat\rho^\perp_S\otimes\1)$ such that
$\la u, (\hat V\hat\rho(0)\hat V)v\ra\neq 0$  
 then  $\ddot p_{ur}(0) <0 $ and there exists $t_0>0$ such that   $\hat\rho_S(t)$ is a mixed state
 for $0<t<t_0$.
\end{proposition}
The formula (\ref{repur}) can be written in a more suggestive form, using the decomposition of $\hat V$ to separate  the contributions of the system and its environment.
The result is 
\beq\label{repur1}
 \ddot p_{ur}(0) = -4\sum_{1\leq j, j'\leq m}\left({\rm tr}_S(\hat\rho_S\hat S_{j'}\hat S_j) - {\rm tr}_S(\hat\rho_S\hat S_{j'}\hat\rho_S\hat S_j)\right){\rm tr}_E(\hat\rho_E\hat E_{j'}\hat E_j)
\eeq
This formula has the following statistical interpretation, introducing the quantum covariance matrices:
\beq\label{repur2}
\Gamma^{(S)}_{j',j} =  {\rm tr}_S(\hat\rho_S\hat S_{j'}\hat S_j) - {\rm tr}_S(\hat\rho_S\hat S_{j'}\hat\rho_S\hat S_j),\;\;\;\;
\Gamma^{(E)}_{j',j} = {\rm tr}_E(\hat\rho_E\hat E_{j'}\hat E_j)
\eeq
so we have
\beq\label{repur3}
 \ddot p_{ur}(0)  = -4 {\rm tr}_{\R^m}(\Gamma^{(S)}(\Gamma^{(E)})^{{\top}}).
\eeq
Then  we  get a short time asymptotic expansion using Taylor formula
$$
p_{ur}(t) = 1 +  \frac{\ddot{p}_{ur}(0)}{2}t^2 + O(t^3),\;\; t\searrow 0.
$$
So if from  formula (\ref{repur3}) we can infer that   $\ddot p_{ur}(0)<0$ then we get that $\hat\rho_S(t)$ is a mixed state for $t>0$ small enough.
The formula (\ref{repur}) is obtained under  rather general conditions. We shall see in the next section  that these conditions are satisfied for the Weyl quadratic model and in particular for quantum Brownian model.\\
Let us give here more general  examples.\\
Assume that  the covariance matrix of $\hat\rho_E(0)$ is diagonal :
${\rm tr}_E(\hat\rho_E(0)\hat E_j\hat E_{j'}) =\delta_{j,j'}d_j$  then we have  from the formula (\ref{repur1})
\beq\label{repur4}
 \ddot p_{ur}(0) = -4\sum_{1\leq j\leq m}\left(\Vert\hat S_j\psi_S\Vert^2 - \la\psi_S,\hat S_j\psi_S\ra^2\right)d_j.
 \eeq
 Hence if there exists $j_0$ such that $d_{j_0}>0$ and $\psi_S$ is not an eigenfunction of $\hat S_{j_0}$
 then  $\ddot p_{ur}(0)<0$.
 
  Let us give more explicit  conditions on Schr\"odinger Hamiltoninans  to satisfy conditions $(GSE)$, $(Ham)_2$, $(S)$  and $(E)$.\\
  Consider  Hamiltonians:
  $\hat H_S =-\triangle_x + V_S(x)$,  $\hat H_E =-\triangle_y+ V_E(y)$, $x\in\R^d$, $y\in\R^N$.\\
  $\hat H_0 = \hat H_S + \hat H_E$, $\hat H = \hat H_0 + V(x,y)$.\\
  The interaction potential $V$ is supposed  such that:
  $\di{V(x,y) =\sum_{1\leq j\leq m}S_j(x)E_j(y)}$,
  Assumming
  \begin{itemize}
  \item $\psi_S\in{\cal S}(\R^d)$, $\rho_E(0)\in {\cal S}(\R^{2N})$
  \item
  $V_S$, $V_E$, $S_j$, $E_j$ are $C^2$ smooth and their derivatives have at most polynomial growth at infinity.
  \item
  $\hat H_0$ and $\hat H$ are self-adjoint (unbounded) operators in $L^2(\R^{(d+N)})$.
  \end{itemize}
Under these conditions the assumptions   of Lemma \ref{anulenv}  are satisfied.
  
 \section{Application to the quadratic case}
  We consider here the Weyl-quadratic case described in  the introduction (all Hamiltonians are  now  supposed  to be quadratic).\\
  We use the following basic property of Weyl quantization in $\R^d$:
 If $A, B$ are Weyl symbols such that $B $ is in the Schwartz  space ${\cal S}(\R^d)$ and $A$ is a polynomial (or like a polynomial)
 then we have
 $$
 {\rm tr}_{L^2(\R^d)}(\hat A\hat B) =(2\pi)^{-d}\int_{\R^{2d}}A(z)B(z)dz.
 $$
 Let us compute   $\ddot p_{ur}(0) $ using formula (\ref{repur2}) and  the Weyl symbols.\\
  Note that $2\pi^{-d}\rho_S(z)dz$ and  $(2\pi)^{-N}\rho_E(u)du $ are  quasi-probabilities laws  denoted  by $\pi_S(z)dz$ and $\pi_E(u)du$  respectively.\\
  Here  we assume that $\hat\rho_S(0)$ is a pure state $\psi_S\in{\cal S}(\R^d)$ and the Weyl symbol (Wigner function)
  is in ${\cal S}(\R^{2N})$.
   \noi
   We can compute
\bea
   {\rm tr}_S(\hat\rho_S\hat S_{j'}\hat\rho_S\hat S_j)  &= &\la\psi_0, \hat S_j\psi_0\ra \la\psi_0, \hat S_{j'}\psi_0\ra\nonumber\\
    & = & (\int_{\R^{2d}}\pi_S(z)S_j(z)dz)(\int_{\R^{2d}}\pi_S(z)S_{j'}(z)dz).
    \eea
    Let us introduce the commutators
    $$
    s_{j'j} = [\hat S_{j'}, \hat S_j],\;\;  e_{j'j} = [\hat E_{j'}, \hat E_j]
    $$
   We have 
   \bea
  {\rm tr}_S(\hat\rho_S\hat S_{j'}\hat S_j)  &= & \int_{\R^{2d}}\pi_S(z)S_{j'}(z)S_j(z)dz + \frac{1}{2i} s_{j'j}\nonumber\\
   {\rm tr}_E(\hat\rho_E\hat E_{j'}\hat E_j)  &= & \int_{\R^{2N}}\pi_E(u)E_{j'}(u)E_j(u)du + \frac{1}{2i} e_{j'j}.
   \eea
   Then we get 
   \bea
   \ddot p_{ur}(0)  = -4\di{\sum_{j,j'}}\left(\int_{\R^{2d}}\pi_S(z)S_{j'}(z)S_j(z)dz  \right.
   \nonumber \\  -  \left.(\int_{\R^{2d}}\pi_S(z)S_j(z)dz)(\int_{\R^{2d}}\pi_S(z)S_{j'}(z)dz)\right)
+  \di{\sum_{j,j'}s_{j'j}e_{j'j}}.
    \eea
Let us introduce  the classical (symmetric) covariance matrices
   $$
   \Sigma^{(S)}_{jj'} = \int_{\R^{2d}}\pi_S(z)S_{j'}(z)S_j(z)dz - 
   (\int_{\R^{2d}}\pi_S(z)S_j(z)dz)(\int_{\R^{2d}}\pi_S(z)S_{j'}(z)dz)
   $$ and 
   $$
   \Sigma^{(E)}_{jj'} = \int_{\R^{2N}}\pi_E(u)E_{j'}(u)E_j(u)du.
   $$
   So we get 
    \beq\label{pur4}
     \ddot p_{ur}(0) = -4{\rm tr}_{\R^{2m}}(\Sigma^{(S)}\Sigma^{(E)})  + \sum_{1\leq j,j' \leq m}s_{j'j}e_{j'j}.
     \eeq
   We remark that in formula (\ref{pur4}) r.h.s   the first term is  classical  and the second term  is a quantum correction 
   $q_c$  which vanishes if the  coupling between the system and its environment  is only in position (or momentum)
   variables ( as it is usually in the literature).

Assume now that $S_j(z) = z_j$ and  $\di{E_j(u) = \sum_{1\leq \ell\leq 2N}G_{j,\ell}u_\ell}$, where $1\leq j\leq 2d$.
$G_{j,\ell}$ is a $2d\times 2N$ matrix with real coefficients.  With these  notations we have for the quantum correction
\beq\label{quantpur}
q_c = \sum_{1\leq j,j' \leq 2d}s_{j'j}e_{j'j} = {\rm tr}_{\R^{2d}}(J_SGJ_EG^{\top})
\eeq
where $J_S, J_E$ are  respectively the matrix of the canonical symplectic form in $\R^{2d}$ and $\R^{2N}$ :
$\sigma_S(z,z') = z\cdot J_Sz'$,  $\sigma_E(u,u') = u\cdot J_Eu'$ (the scalar products are denoted by  a  dot. \\
For  the "classical part " we have 
\beq
-4{\rm tr}_{\R^{2m}}(\Sigma^{(S)}\Sigma^{(E)}) = -4{\rm  tr}(G^{\top}{\rm Cov}_{\rho_S}G{\rm Cov}_{\rho_E})
\eeq
where $G^{\top}$ is the transposed of the matrix $G$.\\
So we have proved the formula:
\beq\label{purfin}
\boxed{ \ddot p_{ur}(0) = -4{\rm  tr}_{\R^{2d}}({\rm Cov}_{\pi_S}G{\rm Cov}_{\pi_E}G^{\top}) +  
{\rm tr}_{\R^{2d}}(J_SGJ_EG^{{\top}})
}
\eeq
where ${\rm Cov}_\pi$ is the covariance matrix for the quasi-probability $\pi$ (the  Wigner function of a pure  state 
$\psi$ is  non negative if and only if $\psi$ is a Gaussian by the Hudson theorem \cite{corob}).
\begin{remark}
From the formula (\ref{purfin})   we get the following inequality
\beq\label{uncert1}
{\rm tr}_{\R^{2d}}(J_SGJ_EG^{{\top}}) \leq 4{\rm  tr}_{\R^{2d}}({\rm Cov}_{\pi_S}G{\rm Cov}_{\pi_E}G^{\top})
\eeq
where $\hat\rho_S$ is a pure state, $\hat\rho_E$ is a centered mixed state, $G$  is a real $2d\times 2N$ matrix.\\
We shall see later that (\ref{uncert1})  countains the uncertainty Heisenberg inequality.
\end{remark}

\noi
In the physicist literature \cite{cale, brpe, flhuro}  the following example is  often considered: a quantum oscillator system is coupled with   a bath of oscillators  in a thermal equilibrium by their  position variables. This is known as the Quantum Brownian Motion model.
So we choose  $d=1$,   $\rho_E = \rho_{\tau,E}$, $\di{V(z, u) = x\cdot\sum_{1\leq j\leq N}c_jy_j}$.\\
$c_1,\cdots, c_N$ are real numbers, $z = (x,\xi)\in\R^2$, $u=(y,\eta)\in\R^{2N}$ and $\rho_E = \rho_{\tau,E}$
where
$$
 \rho_{\tau,E}(u) =  (2\tau)^{N}\exp\left(-\tau\sum_{1\leq j\leq 2N}u_j^2\right)
$$
$\tau$ is related to the temperature  $T$ by the formula $\tau = \tanh(T^{-1})$ (taking the Boltzmann constant  $k=1$).
 So we have $0 <\tau < 1$.\\
 With these parameters we get
\beq\label{purQBM}
 \boxed {\ddot{p}_{ur}(0) = -\frac{2}{\tau}(c_1^2 +\cdots +c_N^2){\rm Cov}_{x,x}(\psi_0)},
\eeq
 where
  ${\rm Cov}_{x,x}(\psi_0)= \int_\R x^2\vert\psi_0(x)\vert^2dx -  (\int_\R x\vert\psi_0(x)\vert^2dx)^2$.
 If the oscillators of the bath have different temperatures $T_1,\cdots, T_N$ then  the result is 
 $$
 \boxed {\ddot{p}_{ur}(0) = - 2(\frac{c_1^2}{\tau_1} +\cdots +\frac{c_N^2}{\tau_N}){\rm Cov}_{x,x}(\psi_0)}
   $$
   where $\tau_j = \tanh(T_j^{-1})$. \\
   In particular we get that $\ddot{p}_{ur}(0)<0$ if $c_{j_0}\neq 0$ for some $j_0$, 
   hence $\hat\rho_S(t)$ is a mixed state for $0<t\leq t_0$, $t_0>0$ small enough.\\
 A more general  bilinear coupling is
 $$
 V(x,\xi;y,\eta) = x\cdot\left(\sum_{1\leq j\leq N} c_jy_j + \sum_{1\leq j\leq N} c_{j+N}\eta_j\right)
  + \xi\cdot\left(\sum_{1\leq j\leq N} d_jy_j + \sum_{1\leq j\leq N} d_{j+N}\eta_j\right)
 $$
 The quantum correction $q_c$ is  obtained by the formula:
 $$
 q_c = 2(d_1c_{N+1} +\cdots d_Nc_{2N}) - 2(c_1d_{N+1} +\cdots c_Nd_{2N})
 $$
 Assume for simplicity that $c_j=0$ for $j\geq N+2$ and $d_j=0$ for $j\geq 2$.
 Then  we find after computations: 
 \bea\label{quc}
 \ddot{p}_{ur}(0) &=& -\frac{2}{\tau}\left((c_1^2 +\cdots +c_N^2+c_{N+1}^2){\rm Cov}_{x,x}(\psi_0) \right.
 \nonumber\\
 & & + \left.2c_1d_1{\rm Cov}_{x,\xi}(\psi_0) + d_1^2{\rm Cov}_{\xi,\xi}(\psi_0)\right) + 
 \underbrace{2c_{N+1}d_1}_{\rm quantum\; correction}
\eea
So we see a  new quantum correction appears  when the coupling mixes  positions and momenta variables.
\begin{remark}
In formula (\ref{quc}) we can see that if  $ {\rm Cov}_{x,\xi}(\psi_0)=0\footnote{the covariance is computed using
the Wigner function of $\psi_0$}$ then $\ddot{p}_{ur}(0)$  can decrease 
or increase with the coefficient $d_1$
compared  to the case $d_1=0$.  \\
Consider the particular case $c_1=\cdots c_N =0$. Then we see easily that $\ddot{p}_{ur}(0)\leq 0$ is equivalent
to  the Heisenberg inequality.
\end{remark}
\begin{remark}
In the above computations we have assumed that the Planck constant  $\hbar$ is one.  We can rewrite formula (\ref{purfin})
including $\hbar$ in Weyl quantization  and we find
\beq
 \ddot p_{ur}(0) = -{4}{\hbar^{-2}}{\rm  tr}_{\R^{2N}}(G^{\top}{\rm Cov}_{\pi_S}G{\rm Cov}_{\pi_E}) +  {\rm tr}_{\R^{2d}}J_SGJ_EG^{\top})
\eeq
As it  is expected the "decoherence time"  becomes  smaller and smaller   when $\hbar\rightarrow 0$. The quantum correction 
has a meaning only when $\hbar$ is not too small.

\end{remark}
\begin{remark}\label{posgauss}
The environment state $\hat\rho_{\tau,E}$ is non-negative  if and only $\tau\leq 1$. It is enough to  prove that for $N=1$. By an  holomorphic extension argument in $\tau$ we can see that $\hat\rho_{\tau,E}$ is negative on  the subspace of $L^2(\R)$ spanned by the odd Hermite functions $\psi_{2k+1}$ where 
$(-\frac{d^2}{dx^2} + x^2)\psi_k = (2k+1)\psi_k$. Let us prove this. 
Denote $\beta = \beta(\tau) = \arg\tanh(\tau)$, 
and $\hat H_{\rm osc} =\frac{1}{2}\left(-\frac{d^2}{dx^2} + x^2\right)$.\\
By the  Mehler formula the Weyl symbol  of  ${\rm e}^{-\beta\hat H_{\rm osc}}$ is
$$
\frac{1}{\cosh(\beta/2)}{\rm e}^{-(\tanh(\beta/2)(x^2+\xi^2)}
$$
The thermal state at temperature $T=\frac{1}{\beta}$ is  defined as 
$$
\hat{\cal T}_\beta =
\frac{ {\rm e}^{-\beta\hat H_{\rm osc}}}{{\rm tr}({\rm e}^{-\beta\hat H_{\rm osc}})}.
$$
Its Weyl symbol is 
$$
{\cal T}_\beta(x,\xi) = \frac{1}{2}\tanh(\beta/2){\rm e}^{-\tanh(\beta/2)(x^2+\xi^2)}.
$$
Let us denote $W_k(x,\xi)$ the Wigner function of $\psi_k$. So we have
\beq
\la\psi_k, {\rm e}^{-\beta\hat H_{\rm osc}}\psi_k\ra = \frac{1}{\cosh(\beta)}
\int_{\R^2}{\rm e}^{-\tanh(\beta)(x^2+\xi^2)}W_k(x,\xi)dxd\xi = {\rm e}^{k+1/2)\beta}
\eeq
Using an  holomorphic extension in the variable $\tau$, for $\tau>1$,  we have
$$
\beta(\tau) = \log\left(\frac{\tau +1}{\tau -1}\right) +\frac{i\pi}{2}
$$
Hence we get
\beq
\la\psi_k,\hat\rho_{\tau,E}\psi_k\ra = \frac{(-1)^k(\tau-1)^k}{2\pi(\tau +1)^{k+1}},\;\; \forall \tau>1.
\eeq
So if $\tau >1$  $\hat\rho_{\tau,E}$ is negative on $\psi_k$  for $k$  odd.
$\square$
\end{remark}
We can apply these results to give a simple proof  that the state the system  $(S)$
 is corollated with the environment $(E)$   for any $t>0$ small enough. 
Here we assume that  both $\hat\rho_S(0)$ and $\hat\rho_E(0)$ are Gaussian pure states. A more general result
 will be given  in Proposition \ref{corr2}. 
\begin{corollary}\label{decohm}
Let $\hat\rho(t) = U(t)\left(\hat\rho_S(0)\otimes\hat\rho_E(0)\right)U^*(t)$ be the time evolution of $(S)\cup(E)$.
Assume  that
  $\ddot p_{ur}(0)\neq 0$ (see formula (\ref{purfin})). Then  there exists $\varepsilon >0$ such that for every $t\in]0, \varepsilon]$ 
we have
\beq\label{dech}
\hat\rho(t) \neq \hat\rho_S(t)\otimes\hat\rho_E(t).
\eeq
\end{corollary}
  {\em Proof}. 
  We shall prove (\ref{dech}) by contradiction. Assume that there  exists a sequence of times $t_n >0$, $\di{\lim_{n\rightarrow +\infty}t_n }=0$
  such that 
  $$
  \hat\rho(t_n) = \hat\rho_S(t_n)\otimes\hat\rho_E(t_n).
  $$
  Let us remark that if such a decomposition exists then  necessarily we have  $\hat\rho_S(t_n) = {\rm tr}_E(\hat\rho(t_n))$
  and   $\hat\rho_E(t_n) = {\rm tr}_S(\hat\rho(t_n))$.\\
  For $t=t_n$ we have
  $$
   {\rm tr}(\hat\rho(t)^2) = {\rm tr}_S(\hat\rho_S(t)^2) {\rm tr}_E(\hat\rho_E(t)^2)
$$
 From the above result  applied to $(S)$ and $(E)$ we have
 $$
 {\rm tr}(\hat\rho(t)^2) = (1 -c_St^2 + O(t^3))(1 -c_Et^2 + O(t^3)) = 1 - (c_S+c_E)t^2 +O(t^3)
 $$
 where all the constants $ c_S, c_E$ are positive.  So we get a contradiction because  ${\rm tr}(\hat\rho(t)^2) =  {\rm tr}(\hat\rho(0)^2) $
 is independent on $t$.
 $\square$\\
 In \cite{eipl} the authors proved  for the quantum Brownian motion  that is  possible to find a Gaussian initial state of the system and a temperature $T$  of the environment such that $\hat\rho(t)$ is not entangled for all times in $[0, +\infty[$.
\begin{remark}
If in the previous corollary $\hat\rho_E(0)$ is a mixed state  and if the linear entropy of $\rho_E(t)$ is increasing in a neighborhood of 0 then $\hat\rho(t) \neq \hat\rho_S(t)\otimes\rho_E(t)$ for $0<t<t_0$  for some $t_0>0$.\\
This is proved by the  method used in the proof of Corollary \ref{decohm}.

\end{remark}
 
 
  \section{The master equation in the Weyl-quadratic case}\label{secquad}
 \subsection{ General quadratic Hamiltonians}
 In this section we find a time dependent  partial differential equation  (often called the  master equation) satisfied by the  reduced density matrix $\hat\rho_S$
  of the open system $(S)$. Moreover this equation can be solved explicitly  using the well known characteristics method.\\
  We extend here  to  any quadratic Hamiltonians with arbitrary bilinear coupling,  
   several results proved in many places  \cite{hpz, hayu, food, flhuro}    for  the quantum Brownian motion model
    with bilinear position coupling.  Our results are inspired by the paper \cite{flhuro} but do not use the path integral methods as  in the papers quoted above. Another difference is that  in  our case the number $N$
    of degrees of freedom
     for the environment is fixed and finite and the number  $d$ of degree of  freedom for the  system is also 
     finite and arbitrary. \\
  There exist  many papers in the physicist literature   concerning  exact or approximated master equation for the 
   quantum Brownian motion model (see the Introduction  and References in  \cite{flhuro}). 
     
  \noi
  Here we assume that $H_S$ and $H_E$ are quadratic Weyl symbols of $\hat H_S$ and $\hat H_E$ respectively.
  We denote $\Phi_0^t$ the Hamiltonian flow generated by $H_0$ and $\Phi^t$ the Hamilton flow generated by 
  $H = H_0+V$.  $V$ is a bilinear   coupling between  the system and the environment :   $V(z,u) =z\cdot Gu$  where $G$ is a linear map from $\R^{2N}$   in $\R^{2d}$ (note that the coupling can mix positions and momenta variables).\\
  Recall that $\R^{2d}$ (resp. $\R^{2N}$) is the classical phase space of the system (resp. the environment).   
   $\Psi^t = \Phi_0^{-t}\Phi^t$ is the interacting flow in the global phase space
  $\R^{2(d+N)}$. The phase space of the global system is identified to the direct sum $\R^{2d}\oplus\R^{2N}$  and in this decomposition the flows are represented by 4 matrix  blocks.
  $$
  \Psi^t = \begin{pmatrix}\Psi_{ii}^t & \Psi_{ie}^t \\ \Psi_{ei}^t & \Psi_{ee}^t\end{pmatrix}
  $$
  The "free" evolution is diagonal
  $$
  \Psi_0^t = \begin{pmatrix}\Phi_S^t &0 \\ 0& \Phi_E^t\end{pmatrix}
  $$
and the interacting classical  Hamiltonian is 
$$
H_I(t,z,u) = V(t,z,u) = z\cdot G(t)u,\;\; {\rm where}\;\; G(t) = (\Phi_S^t)^{\top}\cdot G\cdot\Phi_E^t.
$$
The classical  interacting evolution is given by the equation
$$
\dot\Psi^t = J\nabla^2_{z,u}V(t)\Psi^t,\;\;{\rm with}\; J =\begin{pmatrix}J_S & 0 \\ 0  & J_E \end{pmatrix},
$$
where $\nabla^2_{z,u}V(t)$ is the Hessian matrix in variables $(z,u)\in\R^{2d}\times\R^{2N}$.\\
So for the block  components of the interacting dynamics we have 
\bea\label{classinter}
\dot\Psi_{ii}^t &=  J_SG(t)\Psi^t_{ei},\;\; \dot\Psi_{ie}^t & = J_SG(t)\Psi^t_{ee} \\
\dot\Psi_{ei}^t &= J_EG(t)^{\top}\Psi^t_{ii},\;\; \dot\Psi_{ee}^t &= J_EG(t)^{\top}\Psi^t_{ie}.
\eea
Because all the Hamiltonians considered here are quadratic (eventually  time dependent) they generate well
defined quantum dynamics in Hilbert spaces $L^2(\R^n)$ where $n=d$ (system $(S)$),
 $n=N$ (environment $(E)$) and  $n=d+N$  (global system $(S)\cup (E)$).\\
 Recall the notations $U(t)={\rm e}^{-it\hat H}$, $U_0(t)={\rm e}^{-it\hat H_0}$
 $U_S(t) = {\rm e}^{-it\hat H_S}$,   $U_E(t) = {\rm e}^{-it\hat H_E}$. 
 We have $U_0(t) = U_S(t)\otimes U_E(t)$ and the quantum interacting dynamics 
 $U_I(t) = U_0(t)^*U(t)$.
 

  At time $t=0$ we assume that $\hat\rho(0) =\hat\rho_S(0)\otimes\hat\rho_E(0)$ where $\hat\rho_S$ (resp. $\hat\rho_E $)
 is a  density matrix in the Hilbert space ${\cal H}_S =L^2(\R^d)$ (resp. in ${\cal H}_E =L^2(\R^N)$.\\
 $\rho$, $\rho_S$, $\rho_E$ are the Weyl symbols (i.e the Weyl-Wigner functions,  with $\hbar=1$, 
 of the corresponding density matrices). \\
 The coefficients of the quadratic forms $H_S, H_E, V$  may be time dependent.
 In this case $U(t)$ means $U(t, 0)$ where $U(t, s)$  is the propagator solving
 $$
 i\frac{\partial}{\partial t}U(t, s) = \hat HU(t,s),\;\; U(s,s) =\1_{\cal H}.
 $$
 It is well known that for quadratic Hamiltonians,  every  mixed state $\hat\rho$ of $(S)\cup (E)$ 
 propagates according the classical evolution  (see \cite{corob} for details)
 \beq
 \hat\rho(t) := U(t) \hat\rho(0)U(t)^* ,\;\; \rho(t, X) = \rho(0, \Phi^{-t}X),\; X\in\R^{2(d+N)}
 \eeq
 Our aim is to compute $\hat\rho_S(t)={\rm tr}_E(\hat\rho(t))$.  The Weyl symbol of $\hat\rho_S(t)$  is given by the following integral
\beq\label{weylint}
 \rho_S(t,z) =  \int_{\R^{2N}}\rho(t, z,u)du
 \eeq
 In particular if $\rho(0)$ is Gaussian in all the variables  $X=(z,u)\in\R^{2d}\times\R^{2N}$ 
 then  $\rho_S(t,z) $
  is Gaussian in $z$.  Nevertheless  a direct  computation on the formula (\ref{weylint}) 
  seems   not easy to perform, so a different strategy will be used.

 \subsection{Gaussian mixed states}
Let us consider a very  useful class  of matrix densities with Gaussian symbols; they are   called Gaussian  mixed states. For the reader convenience we recall here  some well known results.
 \begin{definition}
 A  density matrix $\hat\rho$ in the Hilbert space $L^2(\R^n)$ is said Gaussian if  its Weyl symbol $\rho$ is a Gaussian
 $\rho^{\Gamma, m}$ where $\Gamma$ is the covariance matrix (positive-definite $2n\times 2n$ matrix)
 and $m\in\R^{2d}$ the mean of $\rho$. So we have
 \beq\label{gaussmat} 
 \rho^{\Gamma, m}(z) = c_\Gamma{\rm e}^{-\frac{1}{2}(z-m)\cdot\Gamma^{-1}(z-m)},\;\; z=(x,\xi)\in\R^{2n}
 \eeq
 where $c_ \gamma = \det(\Gamma)^{-1/2}$.
  \end{definition}
 Gaussian density matrices are parametrized  by their means  $m$ and their  covariance matrices $\Gamma$
   where we have, as usual
   $$
   m =\int_{\R^{2n}}z\rho^{\Gamma, m}(z)dz,\;\; \Gamma_{j,k} = \int_{\R^{2n}}z_jz_k\rho^{\Gamma, m}(z)dz.
   $$
  As we have seen in Remark \ref{posgauss},  some condition is needed  on the covariance matrix $\Gamma$
  such that (\ref{gaussmat}) defines a density matrix (i.e a non negative operator).
 This condition is a version of the Heisenberg   uncertainty principle (see a proof in the Appendix).
  Here $J$ is the matrix of the symplectic for on $\R^n\times\R^n$  
  and $X=(x,\xi)\in\R^{2n}$.
  \begin{proposition}\label{gausstat}
    $ \rho^{\Gamma, m}(X)$ defines a density matrix if and only if $\Gamma + i\frac{J}{2} \geq 0$, or equivalently,
   if and only if the symplectic eigenvalues of $\Gamma$ are greater than $\frac{1}{2}$.\\
   Moreover, up to a conjugation by a unitary metaplectic transform in $L^2(\R^n)$,  $ \rho^{\Gamma, 0}$
   is a product of one degree of freedom thermal states ${\cal T}_\beta$ (Remark \ref{posgauss}). \\
    $ \rho^{\Gamma, m}(X)$  determines a pure state if and only if    $2\Gamma$ is positive and symplectic
    or equivalently $2\Gamma = FF^\top$ where $F$ is a linear symplectic transformation.
     \end{proposition}
  From these results we can compute  the purity (hence the linear entropy) and the von Neumann entropy of Gaussian states.\\
  For the purity we have the straightforward computation:
\beq
  p_{ur}(\hat\rho^{\Gamma, m}) = (2\pi)^{-n}\int_{\R^{2d}}\rho^{\Gamma, m}(z)^2dz = 
 2^{-n}({\rm det}\Gamma)^{-1/2}.
\eeq
Concerning the von Neumann entropy we begin by   the computation  for thermal states with $n=1$
and temperature $T=\frac{1}{\beta}$.  
For simplicity we compute with the Neper logarithm  ``ln''.  We have
$$ 
\hat\rho_\tau = Z(\beta)^{-1}{\rm e}^{-\beta\hat H_{osc}}
$$
where $\tau = \tanh(\beta/2)$ and 
${\rm ln}({\rm e}^{-\beta\hat H_{osc}}) = -\beta\hat H_{osc}$.  Then  we have   
$$
-{\rm tr}(\hat\rho_\tau{\rm ln}\rho_\tau) = \frac{\beta}{ Z(\beta)}({\rm tr}(\hat H_{osc}{\rm e}^{-\beta\hat H_{osc}}) + {\rm ln}(Z(\beta))
$$
Recall that $Z(\beta) =  \frac{1}{2\sinh(\beta/2)}$. So we compute
$$
{\rm tr}(\hat H_{osc}{\rm e}^{-\beta\hat H_{osc}}) = -\frac{\partial}{\partial\beta}{\rm tr}({\rm e}^{-\beta\hat H_{osc}})
 = \frac{\cosh\beta}{2(\sinh\beta)^2}
 $$
and we get
 $$
{\rm S}_{BN}(\rho_\tau)  = \frac{\beta}{2\tanh(\beta/2)} + {\rm ln}({\rm e}^{\beta/2} - {\rm e}^{-\beta/2}).
 $$
With the parameter $\tau$ we get finally the formula:
 \beq\label{entgauss}
 S_{BN}(\hat\rho_\tau) = \frac{1}{{\rm ln}2}\left(\frac{1-\tau}{2\tau}{\rm ln}\left(\frac{1+\tau}{1-\tau}\right)
  - {\rm ln}\left(\frac{2\tau}{1+\tau}\right)\right)
 \eeq
 The parameter $\tau$ is related  with the linear entropy:
 \beq\label{linentrop}
 S_\ell(\hat\rho_\tau) =  1 - \frac{Z(2\beta)}{Z(\beta)^2} = 1 - \tau,
 \eeq
  These formulas were obtained in \cite{ag}.\\
 Let us denote ${\cal E}(\tau)$ the r.h.s in (\ref{entgauss}). Using Proposition \ref{gausstat}  and additivity  of the von  Neumann  entropy, we  get the following
  formula for a  general Gaussian state  with covariance matrix $\Gamma$:
  \beq
  S_{BN}(\hat\rho^{\Gamma, m}) = \sum_{1\leq j\leq n}{\cal E}(\tau_j)
  \eeq
   where $\tau_1, \cdots, \tau_n$ are the positive symplectic eigenvalues of $\frac{\Gamma^{-1}}{2}$ (see Appendix for more details).

\subsection{Time evolution  of reduced mixed states}

Here we state and prove the main results of this section. It is  convenient to work in the interacting setting. 
Recall that we have
\bea
\hat\rho^{(I)}_S(t) &= & U^*_S(t)\hat\rho_S(0)U_S(t)),\\
\rho^{(I)}_S(t,z) &= &  \rho_S(t, \Phi_S^tz)
\eea
\begin{theorem}\label{masthm}     
Let  $t_c>0$  be the largest  time $t_c$ such that $\Psi_{ii}^t$ is invertible for every $t\in[0, t_c[$.
Assume that the  environment density matrix   $\hat\rho_E(0)$ is a Gaussian with  means $m_E$.  
Then there exist
two  time-dependent  $2d\times 2d$ matrices $A^{(I)}(t)$,   $B^{(I)}(t)$,   
 and a time dependent vector $v^{(I)}(t)\in\R^{2d}$  such that  for $t\in [0, t_c[$ and for every $\rho_S(0)$ 
 in  ${\cal S}(\R^{2d})$, 
the Weyl symbol  $\rho_S^{(I)}(t,z)$ of the interacting evolution $\hat \rho_S^{(I)}(t)$  of the system $(S)$ 
satisfies   the following  master equation (Fokker-Planck type equation):
\bea\label{master1}
\frac{\partial}{\partial t}\rho_S^{(I)}(t,z) &= &(A^{(I)}(t)\nabla_z)\cdot z\rho_S^{(I)}(t,z) +  (B^{(I)}(t)\nabla_z)\cdot \nabla_z\rho_S^{(I)}(t,z) \nonumber \\
 & & + v^{(I)}(t)\cdot\nabla_z\rho_S^{(I)}(t,z).
\eea
Moreover we have the following formula to compute $A^{(I)}(t)$,   $B^{(I)}(t)$ 
\bea\label{FAB}
\label{FA} A^{(I)}(t)^{\top} &=& -J_SG(t)\Psi^t_{ei}(\Psi^t_{ii})^{-1} \\
\label{FB} B^{(I)}(t) &=& \frac{L(t) + L(t)^{\top}}{2},\;\; {\rm where} \\
L(t) &= &J_S\cdot G(t)\left(\Psi^t_{ee} -\Psi_{ei}^t(\Psi_{ii}^t)^{-1}\Psi_{ie}^t\right) 
{\rm Cov}_{\rho_E}(\Psi^t_{ie})^{\top} \\
v^{(I)}(t) & = & -J_SG(t)\Psi_{ee}^tm_E
\eea
\end{theorem}
{\em Proof}.  
 In   this  proof (and only here)  we shall erase the upper index $(I)$ for the interacting dynamics.\\
  Taking the partial trace  in the equation (\ref{LN}) we have
  $$
  \dot\rho_S(t,z) = \int_{\R^{2N}}\{V(t), \rho(t)\}(z,u)du
  $$
  For simplicity we shall assume that $m_E=0$. It is not difficult to take this term into account.\\
  The Poisson bracket is in the variables $(z,u)$  but due to integration in $u$ we have only to consider the
  Poisson bracket in $z$. Hence we get
  \beq\label{derint}
     \dot\rho_S(t,z) =  -\int_{\R^{2N}}J_SG(t)u\cdot\nabla_z\rho(t,z,u)du
     \eeq 
  where $\rho(t,z,u) = \rho(0, \Psi^{-t}(z,u))$. \\
  Denote by $\tilde f(\zeta)$ the Fourier transform of $f$ in the variable $z$. Then we have
  $$
  \tilde{\dot\rho}_S(t,\zeta) = i\zeta\cdot\int_{\R^{2(d+N)}}J_SG(t)u\rho(0, \Psi^{-t}(z,u){\rm e}^{-iz\cdot\zeta}dzdu
$$ 
Now let us perform the symplectic change of variable
$(z',u') = \Psi^{-t}(z,u)$. Then we get, using the splitting assumption at $t=0$, 
\beq
 \tilde{\dot\rho}_S(t,\zeta)  = i\zeta\cdot\int_{\R^{2(d+N)}}dz'du'J_SG(t)(\Psi^t_{ei}z' + \Psi^t_{ee}u')\rho_S(z')
 \rho_E(u'){\rm e}^{-i\zeta\cdot(\Psi^t_{ii}z' + \Psi^t_{ie}u')}.
 \eeq
 Let us denote $\varphi =  \Psi^t_{ii}z' + \Psi^t_{ie}u'$ and using  the equality
 $$
 i(\Psi^t_{ii})^{-1}\nabla_\zeta{\rm e}^{-i\zeta\cdot\varphi} = (z' +(\Psi^t_{ii})^{-1}\Psi_{ie}^tu'){\rm e}^{-i\zeta\cdot\varphi}
 $$
 we get
 \bea
  \tilde{\dot\rho}_S(t,\zeta)  & = &  -\zeta\cdot J_SG(t)\Psi^t_{ei}(\Psi^t_{ii})^{-1}\nabla_\zeta\tilde{\rho}_S(t,\zeta) 
  \nonumber\\
   &  &  - i\zeta\cdot\int_{\R^{2(d+N)}}dz'du'J_SG(t)\Psi^t_{ei}(\Psi^t_{ii})^{-1}\Psi_{ie}^t(u')\rho_E(u')\rho_S(z'){\rm e}^{-i\zeta\cdot\varphi} \nonumber \\
   & & +  i\zeta\cdot\int_{\R^{2(d+N)}}dz'du'J_SG(t)(\Psi_{ee}^tu')\rho_E(u')\rho_S(z'){\rm e}^{-i\zeta\cdot\varphi}
    \eea
To absorb  the linear terms in $u'$ we use that $\rho_E $ is a Gaussian, $\rho_E= c_\Lambda{\rm e}^{-1/2u\cdot\Lambda u}$, 
where $\Lambda$ is positive-definite $2N\times 2N$ matrix and $c_\Lambda$ a normalization constant.
So we have $\Lambda^{-1}\nabla_u\rho_E(u) = u\rho_E(u)$, and   integrating by parts we have
\bea\label{masterf}
  \dot{\tilde\rho}_S(t,\zeta) = - \zeta\cdot A^{(I)}(t)^{\top}\nabla_\zeta\tilde\rho_S(t,\zeta)  - 
  (\zeta\cdot L(t)\zeta)\tilde\rho_S(t,\zeta)  .
  \eea
  We get (\ref{master1})  by inverse Fourier transform.\\
  $\square$\\
  We can deduce a master equation for the  state $\hat\rho_S(t)$ of the system $(S)$:
  \begin{corollary}
  With the notations of Theorem \ref{masthm},  the reduced density matrix for the system  satisfies the following
   master equation
   \bea\label{master2}
\frac{\partial}{\partial t}\rho_S(t,z) &=& \{H_S, \rho_S(t)\}(z)  + (A(t)\nabla_z)\cdot z\rho_S(t,z) +  (B(t)\nabla_z)\cdot \nabla_z\rho_S(t,z) \nonumber\\ & &  + v(t)\cdot\nabla_z\rho_S(t,z), 
\;\;\; {\rm for}\;\; t\in[0, t_c[.
\eea
 where $A(t)$, $B(t)$ and $v(t)$  are given by formulas  (\ref{FAB}) with $\Psi^t = \Phi^t$ (replacing the interacting dynamics by the complete dynamics and  $G(t)$ by $G(0) = G$).
   \end{corollary}
   {\em Proof}.   Recall that $\rho(t, z) = \rho^{(I)}(t, \Phi_S^{-t}(z))$\footnote{If the Hamiltonians are time dependent,
   we have to replace $ \Phi_S^{-t}$ by  $\Phi_S^{(0,t)}$}. So  the change  of variables
    $z\mapsto\Phi_S^{-t}z$ 
   gives easily the result.
   $\square$
  \begin{remark}
   $\rho_S(t,z)$  is of course   well defined for every time $t\in\R$ but the coefficients of the master equation 
   (\ref{master1}) may have singularity at $t=t_c$ as we shall see in examples.  We shall give below a physical interpretation of $t_c$.\\
   A consequence is that $\hat\rho_S(0)\mapsto\hat\rho_S(t)$  is not always a  group of operators.
  \end{remark}
  Let us give a lower bound for the critical time $t_c= \inf\{t>0,\; \det(\Phi^t_{ii}) =0\}$.\\
 Denote $\gamma = \Vert G\Vert$ (it is a measure of the strength of the interaction)
  and $f(t) =\Vert\Phi_S^t\Vert + \Vert\Phi_E^t\Vert$. 
  \begin{proposition}
  If $\di{\sup_{t\geq 0} f(t) < +\infty}$  then there exists $c>0$ such that $t_c \geq  \frac{c}{\gamma}$.\\
  If  there exist $C, \delta >0$ such that $f(t) \leq C{\rm e}^{\delta t}$ for every $t>0$ then there exists $C_1\in\R$
  such that $t_c \geq \frac{1}{\delta}\log(\frac{1}{\gamma}) +C_1$. 
  The constants $\delta$ and $C_1$ are independant of $\gamma$.
  \end{proposition}
  {\em Proof}. 
  It is enough to work in the interaction representation. 
  Using  interacting time evolution     $\Psi^t$  for the total classical system  $(S)\cup (E)$, 
   we get
  $$
  \Vert\Psi^t -\1\Vert \leq \gamma \int_0^tf(s) ds + \gamma\int_0^tf(s)\Vert\Psi^s -\1\Vert ds
  $$
  Denote $F(t) = \int_0^tf(t) ds$. Using the Gronwall Lemma and integrating  by parts we get
  \beq\label{grinq}
   \Vert\Psi^t -\1\Vert \leq \gamma F(t) + \frac{\gamma^3}{6}F(t)^3{\rm e}^{\gamma F(t)}
   \eeq
   From the inequality (\ref{grinq}) we easily get the Proposition.$\square$
  \begin{remark}\label{moment}
  The coefficients of the equation (\ref{master1}) are related with the first and  second moments of  the  reduced density
  matrix $\rho_S(t)$. Let us denote
  $$
  m_j^{(I)}(t) = \int_{\R^{2d}}z_j \rho_S^{(I)}(t,z)dz,\;\;\; \mu_{jk}^{(I)}(t) =   \int_{\R^{2d}}z_jz_k \rho_S^{(I)}(t,z)dz.
  $$
  From  (\ref{master1}) we get
 \beq\label{momt1}
  \dot m^{(I)}(t)   = -(A^{(I)}(t)^{\top}m^{(I)}(t),\;\; {\rm where}\;\; m^{(I)}(t)=(m_1^{(I)}(t),\cdots, m_{2d}^{(I)}(t))
  \eeq
  Computing directly from (\ref{derint}) we get 
  \beq
   \dot m_j^{(I)}(t)  = -\int_{\R^{2d+N}}z_jJ_SG(t)u\cdot\nabla_{z}\rho(t,z,u)dudz  = (J_SG(t)\Psi_{ei}^t m^{(I)}(0))_j.
   \eeq
 So using (\ref{classinter}) we get
 \beq\label{momt2}
  m^{(I)}(t)   = \Psi_{ii}^tm^{(I)}(0) .
  \eeq
From (\ref{momt1}) and (\ref{momt2}) we get  again (\ref{FA}).\\
Computations of  the second moments, using equation  (\ref{master1}),  gives
\beq
\dot\mu^{(I)}(t) =  2B^{(I)}(t)  -(A^{(I),\top}(t)\mu^{(I)}(t) + \mu^{(I)}(t)A^{(I), \top}(t)).
\eeq
As above we can also compute directly using  (\ref{derint})  and (\ref{classinter}) 
\beq
\dot\mu^{(I)}(t)  = \Psi_{ii}^t\mu^{(I)}(0)(\dot\Psi_{ii}^t)^{\top} + \Psi_{ie}^t\mu_E(0)(\dot\Psi_{ie}^t)^{\top}.
\eeq
Using  that  $\mu^{(I)}(t)$ is symmetric we get
\bea
\dot\mu^{(I)}(t)  &=& \frac{d}{2dt}\left(\Psi_{ii}^t\mu^{(I)}(0)(\Psi_{ii}^t)^{\top} + \Psi_{ie}^t\mu_E(0)(\Psi_{ie}^t)^{\top}\right)\\
\mu^{(I)}(t)  &=& \frac{1}{2}\left(\Psi_{ii}^t\mu^{(I)}(0)(\Psi_{ii}^t)^{\top} + \Psi_{ie}^t\mu_E(0)(\Psi_{ie}^t)^{\top}\right).
\eea
where $\mu_E(0)$  is the second moments matrix of $\rho_E(0)$.
\end{remark}

 \begin{remark}
 We can get again the formula (\ref{purfin})  from (\ref{master1}). We have $\dot\rho_S(0) = 0$
  and 
  $$
  \ddot\rho_S(0) = \dot A^{(I)}(0)\nabla_z\rho_S\cdot z\rho_S + \dot B^{(I)}(0)\nabla_z\rho_S\cdot\nabla_z\rho_S.
  $$
  So we get 
   $$
  \dot A^{(I)}(0) = -J_SGJ_EG^{\top},\;\; \dot B^{(I)}(0) =  -J_SG{\rm Cov}_{\rho_E}G^{\top}J_S
   $$
and 
 $$
 {\rm tr}_S(\ddot{\hat\rho}_S(0)\hat\rho_S(0)) =   \frac{1}{2}{\rm tr}_{\R^{2d}}(\dot{A}^{(I)}(0)) + 
 (2\pi)^{-d}\int_{\R^{2d}}\dot{B}^{(I)}(0)\nabla_z\rho_S\cdot\nabla_z\rho_S dz,
 $$
 which co\"{\i}ncides with formula (\ref{purfin})  in this particular case.
 \end{remark}

 \begin{remark}
Suppose that the initial state of the total  system $(S)\cup (E)$ is  Gaussian: \\
$\rho(0,z,u) = c_\Gamma{\rm e}^{-(1/2)\Gamma^{-1}(z,u)\cdot(z,u)}$, $c_\Gamma>0$ is a normalization constant. 
 Then we have the following  direct computation for $\rho_S(t, z)$. \\
 We get first the Fourier transform:
 $$
 \tilde\rho_S(t,\zeta) = {\rm e}^{-(1/2)\Gamma(t)(\zeta, 0)\cdot(\zeta, 0)}
 $$
where $\Gamma(t) =(\Phi^{t})\Gamma(\Phi^{t})^\top$.  Using inverse Fourier transform we have 
 \beq
  \rho_S(t, z) = c_{\Gamma,t}{\rm e}^{-(1/2)\Gamma_S^{-1}(t)z\cdot z}
  \eeq
where $\Gamma_S(t)$ is  the matrix of the positive-definite quadratic form $\zeta\mapsto \Gamma(t)(\zeta,0)\cdot(\zeta,0)$  on $\R^{2d}$  and  
$ c_{\Gamma,t} =  (2\pi)^{-d}\det^{1/2}\Gamma_t $. 

We can  see on this example what is the meaning of  critical times $t_c$.  \\
Assume that $\Gamma  = \Gamma_S \oplus\Gamma_E$ where  $\Gamma_{S,E}$ are 
positive-definite quadratic forms on $\R^{2d}$ respectively $\R^{2N}$ and 
let us introduce the following quadratic forms on $\R^{2d}$.\\
$$
{\cal Q}_t (\zeta,\zeta) = 
\Gamma((\Phi_{ii}^t)^{\top}\zeta, (\Phi_{ei}^t)^{\top}\zeta)\cdot((\Phi_{ii}^t)^{\top}\zeta, (\Phi_{ei}^t)^{\top}\zeta)
$$
We have 
$ {\cal Q}_0(\zeta,\zeta)  = \Gamma_S\zeta\cdot\zeta$   and  
\beq\label{escape}
{\cal Q}_{t} (\zeta,\zeta) =  \Gamma_S^(\Phi_{ii}^t)^{\top}\zeta\cdot(\Phi_{ii}^t)^{\top}\zeta  
 + \Gamma_E(\Phi_{ei}^{t})^{\top}\zeta\cdot(\Phi_{ei}^{t})^{\top}\zeta
\eeq
We see from (\ref{escape}) that the initial state $\hat\rho_S(0)$  cannot be recovered from its
evolution at time $t_c$: only the restriction of ${\cal Q}_{0}$ to $(\ker\Phi_{ii})^\perp$
 is recovered  from ${\cal Q}_{t_c}$. The physical interpretation is that    a part of information contained in $\hat\rho_S(t_c)$  has escaped in the environment represented here by $\Gamma_E$. 
  \end{remark}
  We shall see now that the master equation (\ref{master1}) can be  easily solved by the characteristics method after a  Fourier transform.
  As in the proof of the Theorem \ref{masthm},   $\tilde\rho_S^{(I)}(t,\zeta)$ denotes the Fourier transform of $\rho_S^{(I)}(t)$.
  \begin{theorem}\label{calfrho}
  Assume that the means of the environment  state is 0. With the notations of the Theorem \ref{masthm} we have:
  \beq\label{calrho1}
  \tilde\rho_S^{(I)}(t,\zeta) = \tilde\rho_S(0,(\Psi_{ii}^t)^\top\zeta)\exp\left(-\frac{1}{2}\zeta\cdot\Theta^{(I)}(t)\zeta\right),\; \forall t\in\R, 
  \eeq
  where $\Theta^{(I)}(t) = \Psi^t_{ie}{\rm Cov}_{\rho_E(0)}(\Psi^t_{ie})^{\top}$.
   \end{theorem}
  \begin{remark}
 The   interpretation of the r.h.s in formula (\ref{calrho1}) is the following: the first factor  is a transport term. 
 The second term is a dissipation term due to the influence of the environment which is 
 controlled by the non negative matrix $\Theta^{(I)}(t)$.\\
Formula (\ref{calfrho}) is an extension of  a formula  obtained in \cite{flhuro} for the Brownian quantum motion model where $\Theta^{(I)}(t)$  is named  the "thermal covariance"  when  the environment is in a thermal equilibrium  state.
     \end{remark}
  {\em Proof.} 
  In the proof of the Theorem \ref{masthm} we have seen that  we have $A^{(I)}(t)^\top = -\dot\Psi_{ii}^{(t,0)}\Psi_{ii}^{(0,t)}$. \\
  Here it is convenient to use the notation $\Psi^{(t,s)}$ for the evolution of $\Psi^t$ at time $t$ starting form initial data at time $s$,
  as  it is usual for non autonomous time dependent evolution equations.\\
 So  the characteristic  system   for the equation (\ref{masterf}) is  the linear differential equation
\beq\label{car}
  \dot\zeta = -( \dot\Psi_{ii}^{(t,0)}\Psi_{ii}^{(0,t)})^\top\zeta, \;\; \zeta(0) =\zeta_0.
\eeq
  It is clear that the solution  of (\ref{car})  is $\zeta(t) = (\Psi_{ii}^{(0,t)})^\top\zeta_0$.  So we have 
  \beq
   \tilde\rho_S^{(I)}(t,\zeta) = \tilde\rho_S(0,(\Psi_{ii}^t)^\top\zeta)\exp\left(\int_0^t b(s,\Psi_{ii}^{(s,t)}\zeta)ds\right)
  \eeq
  for $0\leq s\leq t < t_c$, where $b(s, \zeta) = \zeta\cdot B^{(I)}(s)\zeta$. \\
  But  an easy computation shows that we have the following relation
  $$
  B^{(I)}(s) = \frac{1}{2}\left(A^{(I)}(t)^{\top}\Theta^{(I)}(t) + \Theta^{(I)}(t)A^{(I)}(t) + \frac{d}{dt} \Theta^{(I)}(t)\right).
  $$
  Integrating by parts gives
  $$
  \int_0^t b(s,(\Psi_{ii}^{(t,s)})^\top\zeta)ds =-\frac{1}{2}\zeta\cdot\Theta^{(I)}(t)\zeta
  $$
  and formula (\ref{calrho1}) follows.\\
  $\square$\\
  
  It is not difficult to  go back to the evolution of $\tilde\rho_S$ using a change of  variable
  and to the time evolution of $\rho_S(t)$ using inverse Fourier transform.
  \begin{corollary}\label{rhoS}
  Under the conditions of Theorem \ref{calfrho} we have the following formula
  \beq\label{exfour}
   \tilde\rho_S(t,\zeta) = \tilde\rho_S(0,(\Phi_{ii}^t)^\top \zeta)\exp\left(-\frac{1}{2}\zeta\cdot\Theta(t)\zeta\right),
  \eeq
  with
 $ \Theta (t) = \Phi_{ie}^t{\rm Cov}_{\rho_E}(\Phi_{ie}^t)^\top$.\\
 In particular if $\hat\rho(0)$ is a Gaussian state with covariance matrix $\Gamma_S(0)$
  then $\hat\rho_S(t)$ is a Gaussian state with the following covariance matrix  $\Gamma_S(t)$
  \beq\label{covt}
  \Gamma_S(t) = \Phi_{ii}^t\Gamma_S(0)(\Phi_{ii}^t)^\top + \Phi_{ie}^t\Gamma_E(0)(\Phi_{ie}^t)^\top,\;\;  \forall t\in\R, 
  \eeq
  where $\Gamma_E(0) = {\rm Cov}_{\rho_E}$ is the  initial covariance of the environment.
  \end{corollary}
  The formula (\ref{exfour}) is related with Remark \ref{moment} where we have computed the second 
  moments matrix  of $\hat\rho^{(I)}_S(t)$.\\
  As far as $\Phi_{ii}^t$ is invertible we see from (\ref{exfour}) that the time evolution of $\hat\rho_S$ is reversible but this is no more true for $t\geq t_c$.\\
  From  formula (\ref{exfour}) we get an explicit representation formula for the reduced density of the system
  as a convolution integral:
  \begin{corollary}
  With the notations of Corollary \ref{rhoS},   we have
  \bea\label{expred}
  \rho^{(I)}_S(t,z) & = &c_S^{(I)}(t) 
  \int_{\R^{2d}}\rho_S(0,\Psi_{ii}^{(0,t)}z')
  \exp\left(-\frac{1}{2}(z-z')\cdot\Theta^{(I)}(t)^{-1}(z-z')\right)dz', \nonumber\\
  \rho_S(t,z) &=& c_S(t)\int_{\R^{2d}}\rho_S(0,\Phi_{ii}^{(0,t)}z')
  \exp\left(-\frac{1}{2}(z-z')\cdot\Theta(t)^{-1}(z-z')\right)dz'.
   \eea
   where $c_S^{(I)}(t) = (2\pi)^{-d/2}\det(\Psi^t_{ii})^{-1}\det\left(\Theta^{(I)}(t)\right)^{-1/2}$  and  \\
    $c_S(t) = (2\pi)^{-d/2}\det(\Phi^t_{ii})^{-1} \det\left(\Theta(t)\right)^{-1/2}$.\\
    If  $\det(\Phi_{ii}^t) =  0$  or if 
   $\det((\Phi_{ie}^t(\Phi_{ie}^t)^\top)= 0$, integrals in (\ref{expred}) are defined as  convolutions of distributions.  
  \end{corollary}
  Formula (\ref{expred})    shows clearly the  damping   influence of the environment on the system
   because $\Theta(t)$ is a positive  matrix under our assumptions.  If $\Theta(t)$  is degenerated then the Fourier transform 
  of $\exp\left(-\frac{1}{2}\zeta\cdot\Theta(t)\zeta\right)$ is a distribution supported in some linear subspace  $V_S$ of  $\R^{2d}$ where the damping takes place. \\ 
 We can deduce an exact  formula for the linear entropy $S_\ell(t)$ if $\hat\rho_S(0)$ is a Gaussian state.\\
  First, using the Plancherel formula we have
  \bea
  S_\ell(t) &= & 1 - (2\pi)^{-d}\int_{\R^{2d}}(\rho_S(t,z))^2dz \nonumber\\
     & = & 1 - (2\pi)^{-3d}\int_{\R^{2d}}\vert\tilde\rho_S(t,\zeta)\vert^2d\zeta \\
 \label{slg}    & = & 1 -(2\pi)^{-d}\int_{\R^{2d}} 
 \exp(-\zeta\cdot\Theta(t)\zeta)\vert\tilde\rho_S(0,(\Phi_{ii}^t)^\top\zeta)\vert^2d\zeta
     \eea
    If $\rho_S(0)$ is a Gaussian state  we have
    \beq
      S_\ell(t) = 1 -\det\left(\Gamma_S(t)\right)^{-1/2}
      \eeq
  where       $\Gamma_S(t)$ is given by  the formula (\ref{covt}).
  \begin{remark}
  We have seen in this section that we can compute  the quantum evolution of a reduced system 
  when the classical evolution of the blocks $\Phi_{ii}^t$  and $\Phi_{ie}^t$ of the total system
  $(S)\cup (E)$ are known. It is hopeless  to get  general  explicit formulas for  these blocks,
    but this is possible in the  particular case of two oscillators (Appendix \ref{HO}). For  a bath of 
     $N$ oscillators,  $1<N<+\infty$
    the problem seems  difficult (see Appendix \ref{QBM}). For  continuous distribution  of oscillators  many (non rigorous)  results  were  obtained 
     concerning the Quantum Brownian Motion model (see references).
     \end{remark}
      
      Let us  close this  section by giving   a simple sufficient condition to get correlations between 
      states of the system $(S)$ and the environment $(E)$.  If $\hat\rho$ is a state of the total system $(S)\cup (E)$
       we say that $(S)$ and $(E)$ are uncorrelated  if  $\hat\rho = \hat\rho_S\otimes\hat\rho_E$.
        Note that  this decomposition is unique and $\hat\rho_{S,E} ={\rm tr}_{E, S}\hat\rho$.\\
        In decoherence theory \cite{eipl}  more difficult notions  are also considered: separability and entanglement.
        $\hat\rho$ is separable if there exists  a decomposition
        $\di{\hat\rho =\sum_{j} p_j\hat{\rho}^S_j\otimes\hat{\rho}^E_j}$ where $\di{\sum_jp_j =1}$, $p_j\geq 0$,
        $ \hat{\rho}^S_j$, $\hat{\rho}^E_j$ are pure states respectively  in  ${\cal H}_S$, ${\cal H}_E$.
          If $\hat\rho$ is not separable it is  said  that  $\hat\rho$ is entangled.
      
        \begin{proposition}\label{corr2}
        Assume that $\hat\rho(0) = \hat\rho_S(0)\otimes\hat\rho_E(0)$ and that
        $\hat\rho_S(0), \hat\rho_E(0)$ are Gaussian states with covariance matrices 
        $\Gamma_S, \Gamma_E$.\\
        If  the following  generic condition is satisfied
       \beq\label{cnent}
        J_S\Gamma_SGJ_E\Gamma_E^{-1} + G \neq 0
       \eeq
        then there exists $\varepsilon >0$  such that   for every  $t\in]0, \varepsilon]$, 
        $\hat\rho(t)\neq\hat\rho_S(t)\otimes\hat\rho_E(t)$, which means that  the reduced system $(S)$ is immediately   correlated  with its environment $(E)$
         when a coupling is switch on.
        \end{proposition}
        {\em Proof}.  The proof is elementary.\\
        We compute  with  the classical  Gaussian state in the interaction representation (as above
        for simplicity we erase the index $(I)$). We have
        $$
        \rho(t,z,u) = \rho_S(0, \Psi_{ii}^{(0,t)}z + \Psi_{ie}^{(0,t)}u)\rho_E(0,\Psi_{ee}^{(0,t)}u + \Psi_{ei}^{(0,t)}z).
        $$
        Using the assumptions, we have
        \beq
        \rho(t,z,u) = K(t,z)L(t,u){\rm e}^{-z\cdot C(t)u}
      \eeq
       where the crossed term is determined by the matrix $C(t)$  satisfying $C(0) =0$.
       Using the  interacting evolution equation we can easily compute the derivative $\dot C(0)$.
       So we find $\dot C(0) = \Gamma_S^{-1}J_SG - GJ_E\Gamma_E^{-1}$ and $\dot C(0)\neq 0$
       if condition (\ref{cnent}) is satisfied. Hence $C(t)\neq 0$ for $t$ small enough and $t\neq 0$
        $\square$
       \begin{remark}
       In \cite{eipl} the authors have proven a stronger entanglement  result for the Quantum Brownian Motion model.
       \end{remark}
 \appendix\section {Coupled Harmonic oscillators}\label{HO}
 More explicit  computations can be done  for systems  with two  coupled one dimensional harmonic oscillators.\\
 $$
 \hat H_S=\frac{1}{2}(-\partial^2_{x^2} +{\omega_S}^2x^2),\;\;\; 
 \hat H_E=\frac{1}{2}(-\partial^2_{y^2} +{\omega_E}^2y^2).
 $$
 We assume that $\omega_S >0$ and $\omega_E$   can be  real positive or purely imaginary ($\omega_E^2 <0$). 
 $\omega^2_E>0$ means a stable environment and   $\omega^2_E<0$ means an unstable environment.
 Unstable environment was considered in the paper \cite{blzu}. Here we revisit  the computations of \cite{blzu}.\\

\noi
 Le us compute the classical flow of the total Hamiltonian
 $$
 H(x,\xi,y,\eta) = \frac{1}{2}(\xi^2 + \eta^2+ {\omega_S}^2x^2 + {\omega_E}^2y^2) +\gamma xy.
 $$
 Let  $M=\begin{pmatrix}\omega_S^2  &\gamma\\ \gamma & \omega_E^2    \end{pmatrix}$ 
 be  the matrix of the quadratic potential  for the total system: $V(x,y) = {\omega_S}^2x^2 + {\omega_E}^2y^2 + \gamma xy$.\\
 The eigenvalues $\lambda_\pm$ of $M$ are
 $$
 \lambda_\pm = \frac{1}{2}\left(\omega_S^2 + \omega_E^2 \pm \sqrt{(\omega_S^2-\omega_E^2)^2 +4\gamma^2}\right)
 $$
 So we have
 $$
  M = P\begin{pmatrix}\lambda_+ & 0\\ 0 & \lambda_-  \end{pmatrix}P^{-1}
  $$ 
  where 
  $P = \begin{pmatrix}\cos\theta & -\sin\theta\\ \sin\theta & \cos\theta  \end{pmatrix}$ and 
  \bea
  \cos\theta = \left(\frac{1}{2}(1 + 
  \frac{\omega_S^2 -\omega_E^2}{(\omega_S^2 -\omega_E^2)^2 +4\gamma^2)^{1/2}})\right)^{1/2} \nonumber\\
  \sin\theta = \left(\frac{1}{2}(1 -
  \frac{\omega_S^2 -\omega_E^2}{(\omega_S^2 -\omega_E^2)^2 +4\gamma^2)^{1/2}})\right)^{1/2} 
   \eea
   Now we compute the matrix of $\Phi^t$ in the canonical basis  of $\R_x\times\R_y\times\R_\xi\times\R_\eta$
   $$
   e_x=\left(\begin{array}{c} 1 \\0\\0\\0\\ \end{array}\right),\; e_y=\left(\begin{array}{c} 0 \\1\\0\\0\\ \end{array}\right),\;
    e_\xi=\left(\begin{array}{c} 0 \\0\\1\\0\\ \end{array}\right), \; 
    e_\eta=\left(\begin{array}{c} 0 \\0\\0\\1\\ \end{array}\right)
   $$
   We have $\Phi^t = \Pi\Phi_{\rm diag}^t\Pi^{-1}$ where
   $$
   \Pi = \begin{pmatrix} P & 0_2\\ 0_2 & P\end{pmatrix},
   $$ 
   and 
   $$
   \Phi_{\rm diag}^t = \begin{pmatrix} \cos(t\lambda_+^{1/2}) & 0 &\lambda_+^{-1/2}\sin(t\lambda_+^{1/2}) &0\\
   0 & \cos(t\lambda_-^{1/2})  &0 & \lambda_-^{-1/2}\sin(t\lambda_-^{1/2})\\
   -\lambda_+^{1/2}\sin(t\lambda_+^{1/2}) & 0 &\cos(t\lambda_+^{1/2}) &0\\
   0 &  -\lambda_-^{1/2}\sin(t\lambda_-^{1/2}) & 0 & \cos(t\lambda_-^{1/2})
     \end{pmatrix}
     $$
     The coefficients  of the matrix $ \Phi^t$ are easily computed  using the formula
      $$
      e_\alpha\cdot\Phi^te_\beta = \Pi^{-1}e_\alpha\cdot\Phi^t_{\rm diag}\Pi^{-1}e_\beta.
      $$
      Let us denote $\Phi^t(j,k)$ the coefficient with raw index $j$ and  column  index $k$ of the matrix $\Phi^t$.
      This matrix has four $2\times 2$ blocks:  $\Phi_{ii}^t, \Phi^t_{ie}, \Phi^t_{ei}\Phi^t_{ee}$. (16 terms).
     So we have $\Phi^t_{ii}(j,k) = \Phi^t(j,k)$ for $j, k\in\{1,2\}$, $\Phi^t_{ie}(j,k) = \Phi^t(j,k)$ with $j\in\{1,2\}$,
     $k\in\{3,4\}$,  $\Phi^t_{ei}(j,k) = \Phi^t(j,k)$ with $j\in\{3,4\}$,
     $k\in\{1,2\}$,  $\Phi^t_{ee}(j,k) = \Phi^t(j,k)$ with $j, k\in\{3,4\}$,
     
            \bea
     \Phi_{ii}^t(1,1)  &= &\cos^2\theta\cos(t\lambda_+^{1/2}) +\sin^2\theta\cos(t\lambda_-^{1/2}) \\
     \Phi_{ii}^t(1,2)  &=& \lambda_+^{-1/2}\cos^2\theta\sin(t\lambda_+^{1/2}) + \lambda_-^{-1/2}\sin^2\theta\sin(t\lambda_-^{1/2}) \\
      \Phi_{ii}^t(2,1)  &= &-\lambda_+^{1/2}\cos^2\theta\sin(t\lambda_+^{1/2}) - \lambda_-^{1/2}\sin^2\theta\sin(t\lambda_-^{1/2}) \\
       \Phi_{ii}^t(2,2)  &=&  \cos^2\theta\cos(t\lambda_+^{1/2})  +\sin^2\theta\cos(t\lambda_-^{1/2})
     \eea
    We have analogous formula for the blocks  $\Phi^t_{ie}$,  $\Phi^t_{ei}$,  $\Phi^t_{ee}$.\\
     \bea
      \Phi_{ie}^t(1,1) &=& \sin\theta\cos\theta(\cos(t\lambda_+^{1/2}) -\cos(t\lambda_-^{1/2})) \\
       \Phi_{ie}^t(1,2) &=& \sin\theta\cos\theta(\lambda_+^{-1/2}\sin(t\lambda_+^{1/2})-\lambda_-^{-1/2}\sin(t\lambda_-^{1/2}))  \\
         \Phi_{ie}^t(2,1) &=&\sin\theta\cos\theta(-\lambda_+^{1/2}\sin(t\lambda_+^{1/2})+\lambda_-^{1/2}\sin(t\lambda_-^{1/2}))\\
       \Phi_{ie}^t(2,2) &=&  \sin\theta\cos\theta( \cos(t\lambda_+^{1/2}) -\cos(t\lambda_-^{1/2}))
\eea
We note that $\Phi^t_{ei}=\Phi^t_{ie}$.
\bea
  \Phi_{ee}^t(1,1) &=& \cos^2\theta\cos(t\lambda_-^{1/2}) +\sin^2\theta\cos(t\lambda_+^{1/2})\\
\Phi_{ee}^t(1,2) &=&   \lambda_-^{-1/2} \cos^2\theta\sin(t\lambda_-^{1/2}) + \lambda_+^{-1/2}\sin^2\theta\sin(t\lambda_+^{1/2})\\
\Phi_{ee}^t(2,1) &=&   -\lambda_-^{1/2}\cos^2\theta\sin(t\lambda_-^{1/2}) -\lambda_+^{1/2}\sin^2\theta\sin(t\lambda_+^{1/2})\\
\Phi_{ee}^t(2,2) &=&  \cos^2\theta\cos(t\lambda_-^{1/2})  +\sin^2\theta\cos(t\lambda_+^{1/2})
\eea
     In particular we have
     \bea\label{deter1}
     \det\left( \Phi_{ii}^t\right) = 1 + 2\sin^2\theta\cos^2\theta \nonumber \\
     \left(\cos(t\lambda_+^{1/2})\cos(t\lambda_-^{1/2})
      + \frac{\lambda_+ + \lambda_-}{2\sqrt{\lambda_+\lambda_-}}\sin(t\lambda_+^{1/2})\sin(t\lambda_-^{1/2}) -1\right)
     \eea
     and 
     \bea\label{deter2}
     \det\left( \Phi_{ii}^t\right) = 1 + 2\frac{\gamma^2}{(\omega_S^2-\omega_E^2)^2 +4\gamma^2}\nonumber\\
     \left(\cos(t\lambda_+^{1/2})\cos(t\lambda_-^{1/2})
      + \frac{\omega_S^2+\omega_E^2}{2\sqrt{\omega_S^2\omega_E^2-\gamma^2}}
      \sin(t\lambda_+^{1/2})\sin(t\lambda_-^{1/2}) -1\right)
      \eea
     From equation (\ref{deter2}) we can compute the critical time $t_c$.  We see here that we  may have $t_c$ finite or infinite,
      depending on the coupling constant $\gamma$ and on the sign of $\omega_E^2$.\\
  1. Assume that $\omega_S^2\omega_E^2 > \gamma^2$ and $\omega_S^2\neq\omega_E^2$.\\
  We have $\lambda_+>\lambda_->0$. We see easily that, for some constant $C$ depending only on 
  $\omega_S, \omega_E$,  we have 
  $$
  \vert\det\left( \Phi_{ii}^t\right) - 1\vert \leq C\gamma,\;\; \forall t\in\R
  $$
 Then if $\gamma < \frac{1}{C}$, $(\Phi_{ii}^t)$ is invertible for every $t\in\R$.\\
 2.      If  $\omega^2_S=\omega^2_E$   and $\omega_S^2>\gamma^2$ we get from (\ref{deter2}) that there
 exists $C>0$ such that   $\Phi_{ii}^t$ is invertible for $0\leq t \leq \frac{C}{\gamma}$.\\
 3. If $\omega_S^2\omega_E^2 < \gamma^2$  then $\lambda_+ > 0 > \lambda_-$.  
 We can find constants $C_1, C_2$ such that if $\gamma{\rm e}^{C_2t} \leq C_1$
  then  $\Phi_{ii}^t$ is invertible.

     \section{On the quantum Brownian motion model} \label{QBM}
     The quantum brownian motion model was introduced in the paper \cite{fokama}. 
     The starting point is a  one degree of  freedom system interacting with a large number $N$
     of harmonic oscillators.
     The total (classical) Hamiltonian  is the following
     \beq
     H(x,\xi;y,\eta) = \frac{\xi^2}{2m} + V(x) +\sum_{1\leq j\leq N}\frac{\eta_j^2}{2m_j} + k_j(y_j-x)^2.
  \eeq
     where $x, \xi$ are the coordinates of the system, $y=(y_1,\cdots, y_N)$ and
     $\eta =(\eta_1,\cdots,\eta_N)$  are the coordinates  for the environment.
     The environment  consists in $N$ harmonic oscillators and the system is connected 
      to each oscillator by a spring with constant $k_j>0$.  This model is a particular  case of  quadratic  systems
   considered  in section \ref{secquad}. 
       if $V(x) = \frac{\omega_S^2}{2}x^2$. The Hamiltonian  can be splitted as
       $H = H_S+H_E +H_I$  where 
          \bea
   H_S &=&\frac{\xi^2}{2m} + V(x) + \frac{1}{2}(\sum_{1\leq j\leq N}k_j)x^2 \\
     H_E &=& \frac{1}{2}\left(\sum_{1\leq j\leq N}\frac{\eta_j^2}{m_j} + k_jy_j^2\right)\\
     H_I &=& -2x(\sum_{1\leq j\leq N} k_jy_j)
     \eea
    The classical evolution for the Hamiltonian is not explicitly given for $N>1$ (we have got explicit formula
    in Appendix \ref{HO} if $N=1$).  It can be seen that the time evolution of the position $x$ 
    of the system satisfies the following  equation \cite{foka}, assuming $m_j=1$ for simplicity:
    \beq\label{inteq}
    m \ddot{x}(t) +\int_0^tK(t-s)\dot{x}(s)ds +V'(x(t)) + K(t)x(0) =F(t)
    \eeq
    where 
    \bea
    K(t) &=& \sum_{1\leq j\leq N}k_j\cos(\sqrt k_jt) \\
    F(t) &=& \sum_{1\leq j\leq N}y_j(0)\cos(\sqrt k_jt) + \eta_j(0)\sin(\sqrt k_jt)
    \eea
   The difficulty here is that equation (\ref{inteq}) is not an ODE because of the integral term. \\
   In \cite{fokama, foka} the authors considered  a large $N$ limit, 
    and a continuous distribution of oscillators,  such that (\ref{inteq})
   is transformed into a stochastic differential equation where the integral term is replaced by the damping term $\gamma\dot x(t)$ where $\gamma$  is a damping constant (Langevin equation).
     
     \section{Gaussian density matrices}
     
     We  shall give  here a proof of Proposition \ref{gausstat}.
    This is 	 consequence of  a particular case of  the following Williamson theorem (see \cite{wil} and  \cite{ar} Appendix 6).
     \begin{theorem}\label{william}
     Let $\Gamma$ be a positive  non  degenerate  linear transformation  in  $\R^{2N}$. Then   there exists a  linear symplectic
      transformation  $S$ and positive real numbers  $\lambda_1, \cdots, \lambda_N$  such that 
      \beq\label{will}
      S^\top\Gamma Se_j =  \lambda_je_j,\;{\rm and} \;   S^\top\Gamma Se_{j+N} =  \lambda_je_{j+N}
      \eeq
       for $1\leq j\leq N$, where $\{e_1,\cdots, e_N,\cdots, e_{2N}\}$
       is the canonical basis  of $\R^{2N}$.
     \end{theorem}
   in \cite{sichsr} the authors gave a simple proof that we recall here.
   Recall the following known lemma
   \begin{lemma}\label{lemA}
   Let  $A$ be a non degenerate  antisymmetric linear mapping in $\R^{2N}$. Then there exists an orthonormal basis 
     $\{v_1, v_1^*,\cdots, v_N, v_N^*\}$  of $\R^{2N}$   and  positive real numbers $\{\nu_1, \cdots, \nu_N\}$, such that 
   $$
   A v_j = \nu_j v_j^*,\;\;  A v_j^* = -\nu_j v_j.
   $$
   \end{lemma}
   {\em Proof of Lemma \ref{lemA} }.  
   We proceed by induction on $N$. This is is obvious for $N=1$.\\
   Assume $N\geq 2$.  Let $\nu_1$  be an eigenvalue of  the symmetric matrix $A^2$,
   $A^2v_1 = \nu_1v_1$, $\Vert v_1\Vert =1$.  We can choose  a vector $v_1^*$  and $a\in\R$
    such that $Av_1 = -av_1^*$ and $\Vert v_1^*\Vert =1$.  Then we have easily that $v_1\cdot v_1^* = 0$ 
     and if $P$
    is the plane spanned by $\{v_1, v_1^*\}$  then $P$ and $P^\perp$ are invariant by $A$.
    So we can  apply the induction assumption to $A$ acting in $P^\perp$ and the Lemma is proved.
    $\square$

    \noi
     {\em Proof of Theorem \ref{william}. }
Consider the  antisymmetric matrix $A = \Gamma^{-1/2}J\Gamma^{-1/2}$. Using Lemma \ref{lemA} we can find
an orthogonal matrix $R$ and a diagonal matrix 
$\Omega = {\rm diag}\{\nu_1,\nu_2, \cdots, \nu_N \}$, $\nu_j>0$, $1\leq j\leq N$  such that
$$
R^\top\Gamma^{-1/2}J\Gamma^{-1/2}R = \begin{pmatrix} 0 & \Omega \\ -\Omega & 0 \end{pmatrix}
$$
Denote $B = \begin{pmatrix} \Omega^{-1/2} & 0 \\ 0 & \Omega^{-1/2} \end{pmatrix}$
 and $S = \Gamma^{-1/2}RB$.  We get easily
 that $S^\top\Gamma S = B^2$ and  $S^\top J S = J$ so the proof of the Theorem \ref{william} follows.
 $\square$
 
   \noi The $N$ real numbers $\lambda_j$  in Theorem \ref{william}  are the symplectic eigenvalues of $\Gamma$.
   Using that $S^\top = -JS^{-1}J$ we see  that 
 $J\Gamma$ is diagonalizable  with eigenvalues  $\{\pm\lambda_j, \;1\leq j\leq N\}$  and that  
 $\{\lambda_1, \cdots, \lambda_N\}$ are the eigenvalues  (with multiplicities) of $\vert J\Gamma\vert = (-J\Gamma^2J)^{1/2}$.

 \noi
{\em Proof of Proposition \ref{gausstat}}
We can assume assume that $m=0$ and we denote $\rho_\Gamma =\rho^{\Gamma, 0}$.\\
We use the symplectic normal form for $\Gamma$ given by Theorem \ref{william}. Let $\hat R(S)$ be the metaplectic  unitary operator  associated with $S$ (see for example \cite{corob}).  Hence  we have
$$
\hat R(S)^*\hat\rho_\Gamma\hat R(S) = \hat\rho_{S^\top\Gamma S} = \hat\rho_{\tau_1}\otimes \hat\rho_{\tau_1}\cdots  \otimes\hat\rho_{\tau_N}, 
$$
where the $\tau_j$ are the symplectic eigenvalues of $\frac{\Gamma^{-1}}{2}$.  So 
$\hat\rho_\Gamma$ is a density matrix if and only if  we have $0\leq \tau_j \leq 1$ for $1\leq j\leq d$ 
(Remark \ref{posgauss}).  This condition  means that  the symplectic eigenvalues of $2\Gamma$ are 
greater than 1  or equivalently  that $2\Gamma + iJ\geq 0$.\\
We have already seen that if $\hat\rho_\Gamma$ is a pure state then we have $2\Gamma = F^\top F$
with $F$ symplectic. Conversely if  $2\Gamma = F^\top F$  then $\hat\rho_\Gamma $ is the Wigner function
of a squeezed state  $\hat R(F)\varphi_0$, $\varphi_0$ being the standard Gaussian (for details see \cite{corob}).

Gaussian states are  thermal states for    positive non degenerate 
  quadratic Hamiltonians  and conversely. This  can be proved as follows.\\ 
  Let  $H(u) = \frac{1}{2}u\cdot \Lambda$  where $\Lambda $ is  a  positive non degenerate  linear transformation
   in $\R^{2N}$.   We consider  symplectic coordinates for  $u$:  $y_j=u\cdot e_j$
    and $\eta_j = u\cdot e_{j+N}$, $1\leq j\leq N$.\\
    Using Theorem \ref{william}  we have
   $H(u) = H_\Delta(Su)$  where $S$ is a symplectic linear transformation  and
   $\di{H_\Delta(u)  = \sum_{1\leq j\leq N}\lambda_ju_j^2}$,  $u_j =(x_j, \xi_j)\in\R^2$.
   \noi
      Applying the metaplectic transformation we have $\hat H= \hat R(S)\hat H_\Delta\hat R(S)^*$ 
      and 
   $$
   {\rm e}^{-\beta\hat H}= \hat R(S){\rm e}^{-\beta\hat H_\Delta}\hat R(S)^*
   $$
   From the Melher formula for the harmonic oscillator the Weyl symbol $W_\Delta$ of 
   ${\rm e}^{-\beta\hat H_\Delta} $ is
   $$
   W_\Delta(y,\eta) = \prod_{1\leq j\leq N}
   \frac{\exp\left(-\tanh(\beta\lambda_j)(y_j^2+\eta_j^2)\right)}{\cosh(\beta\lambda_j)}
   $$
   so  the Weyl symbol $W$ of  ${\rm e}^{-\beta\hat H}$ is given by
   $$
   W(y,\eta) = W_\Delta(S^{-1}(y,\eta)).
   $$
 This proves that   $\frac{ {\rm e}^{-\beta\hat H}}{{\rm tr} {\rm e}^{-\beta\hat H}}$ is a Gaussian state.
 More  explicit results are given in \cite{de, ho2}.


\rule[0pt]{\hsize}{0.4pt}

\noi
D\'epartement de Math\'ematiques,  Laboratoire Jean Leray, CNRS-UMR 6629\\
Universit\'e de Nantes, 2 rue de la Houssini\`ere,  F-44322 NANTES Cedex 03, France\\
E-mail adress: \texttt{didier.robert@univ-nantes.fr}

\end{document}